\documentclass[fleqn,usenatbib]{mnras}

\usepackage{mathptmx}

\usepackage[T1]{fontenc}

\usepackage[dvipsnames]{xcolor}

\DeclareRobustCommand{\VAN}[3]{#2}
\let\VANthebibliography\thebibliography
\def\thebibliography{\DeclareRobustCommand{\VAN}[3]{##3}\VANthebibliography}

\usepackage{graphicx}   
\usepackage{amsmath}    
\usepackage{amssymb}    

\definecolor{ao(english)}{rgb}{0.0, 0.5, 0.0}

\colorlet{gwpurple}{blue!40!red}

\title[Funnel properties in radiation GRMHD]{Radiation GRMHD simulations of M87: funnel properties and prospects for gap acceleration}

\author[Yao et al.]{
Philippe Z. Yao,$^{1,2}$\thanks{zhiyu.yao@colorado.edu}
Jason Dexter,$^{1,2}$\thanks{jason.dexter@colorado.edu}
Alexander Y. Chen$^{2}$,
Benjamin R. Ryan,$^{3}$
\newauthor \,George N. Wong$^{4}$
\\
$^{1}$Department of Astrophysical and Planetary Sciences, University of Colorado, 391 UCB, Boulder, CO 80309-0391, USA\\
$^{2}$JILA, University of Colorado and National Institute of Standards and Technology, 440 UCB, Boulder, CO 80309-0440, USA\\
$^{3}$CCS-2, Los Alamos National Laboratory, P.O. Box 1663, Los Alamos, NM 87545, USA\\
$^{4}$Department of Physics, University of Illinois, 1110 West Green Street, Urbana, IL 61801, USA
}

\date{Accepted XXX. Received YYY; in original form ZZZ}

\pubyear{2021}

\begin{document}
\label{firstpage}
\pagerange{\pageref{firstpage}--\pageref{lastpage}}
\maketitle

\begin{abstract}
We use the public code \textsc{ebhlight} to carry out 3D radiative general relativistic magnetohydrodynamics (GRMHD) simulations of accretion onto the supermassive black hole in M87. The simulations self-consistently evolve a frequency-dependent Monte Carlo description of the radiation field produced by the accretion flow. We explore two limits of accumulated magnetic flux at the black hole (SANE and MAD), each coupled to several sub-grid prescriptions for electron heating that are motivated by models of turbulence and magnetic reconnection. We present convergence studies for the radiation field and study its properties. We find that the near-horizon photon energy density is an order of magnitude higher than is predicted by simple isotropic estimates from the observed luminosity. The radially dependent photon momentum distribution is anisotropic and can be modeled by a set of point-sources near the equatorial plane. We draw properties of the radiation and magnetic field from the simulation and feed them into an analytic model of gap acceleration to estimate the very high energy (VHE) gamma-ray luminosity from the magnetized jet funnel, assuming that a gap is able to form. We find luminosities of $\rm \sim 10^{41} \,  erg \, s^{-1}$ for MAD models and $\rm \sim 2\times 10^{40} \, erg \, s^{-1}$ for SANE models, which are comparable to measurements of M87's VHE flares. The time-dependence seen in our calculations is insufficient to explain the flaring behavior. Our results provide a step towards bridging theoretical models of near-horizon properties seen in black hole images with the VHE activity of M87.
\end{abstract}





\section{Introduction}

The relativistic jet launched from the center of the elliptical galaxy M87 radiates across the radio to X-ray bands \citep[e.g.,][]{biretta1999,junor1999,dimatteo2003,walker2018}. Radio VLBI observations show that the jet core originates near the black hole event horizon \citep{hada2011}. The jet is also a source of rapidly variable, very high energy TeV emission \citep{2006Sci...314.1424A}. The variability time scale can be as short as 1--2 days \citep{abramowski2012}, 
which corresponds to fewer than 10 light-crossing times of the event horizon of a black hole with mass $M_\mathrm{BH}\approx 6\times 10^9M_\odot$. 

The physical origin of the flaring remains uncertain. During the 2008 flaring events, radio brightening and subsequent enhanced X-ray flux from the core region was detected \citep{2009Sci...325..444A}. The multi-wavelength correlations together with the rapid variability suggest that the VHE $\gamma$-ray emission may originate from the jet base, close to the event horizon.

The jet is thought to be launched from the accretion flow onto M87's central, supermassive black hole \citep[e.g.,][]{reynolds1996,dimatteo2003}. The accretion flow is expected to be hot, geometrically thick, and optically thin at the low luminosity of M87, $L/L_{\rm Edd} \lesssim 10^{-5}$ \citep{kuo2014}, where $L_{\rm Edd} \equiv 4\pi GMm_pc/\sigma_T$ is the Eddington luminosity. A theoretical description of magnetized accretion \citep{mri} and jet launching via the Blandford--Znajek mechanism \citep{blandfordznajek} can be captured in global, three dimensional, general relativistic magnetohydrodynamic (GRMHD) simulations \citep[e.g.,][]{gammie2003,devilliers2003}. Such simulations can be used to interpret observations of near horizon radiation from M87 \citep[e.g.,][]{moscibrodzka2011,dexter2012,moscibrodzka2016,ryan2018,EHTPaperV,chael2019,davelaar2019}, especially in the context of the resolved structure at $1.3$mm \citep{doeleman2012,akiyama2015,EHTPaperIV,EHTPaperVI,wielgus2020}.

Unlike in the case of the Galactic center \citep{dibi2012,drappeau2013}, radiative cooling of relativistic electrons by inverse Compton scattering and synchrotron radiation is likely important at the comparatively higher mass accretion rate of the M87 black hole system \citep[e.g.,][]{moscibrodzka2011}. Cooling has been incorporated in GRMHD simulations of accretion onto the M87 black hole
using both an M1 moment scheme \citep{chael2019} and a full frequency- and angle-dependent Monte Carlo treatment of the radiation field \citep{ryan2018}. In both cases, the electrons are evolved as a separate fluid \citep{ressler2015}, whose heating is modeled using sub-grid prescriptions based on kinetic physics \citep[e.g.,][]{howes2010,rowan2017}. Using a specific electron heating model, radiation GRMHD simulations can self-consistently evolve the radiation field and model the observed spectrum due to synchrotron emission and Compton scattering.

The TeV VHE emission, on the other hand, is produced by electrons accelerated to very high energies ($\gamma_e \gtrsim 10^6$). One proposed acceleration site could be near the event horizon, where a charge-starved ``gap'' region could form within the magnetically dominated jet near the pole (the ``funnel'') \citep{2011ApJ...730..123L}. Inside this gap region, particles would be accelerated by an unscreened electric field $E_\parallel$ (parallel to the magnetic field in the jet) to Lorentz factors of $\gamma_e > 10^6$ and emit gamma-rays through inverse-Compton scattering with the photons produced in the disk. Some of these gamma-rays convert to pairs to sustain the electric current in the jet funnel, while the rest escape to infinity and produce the observed emission. 
Since $E_\parallel$ is explicitly zero in ideal GRMHD simulations, parallel acceleration cannot be captured self-consistently. Simulations that treat $E_\parallel \neq 0$ would have to account for highly nonthermal particle distributions.
 Recently this process was studied through first-principles local and global Particle-in-Cell (PIC) kinetic simulations \citep[e.g.,][]{levinson2018,chen2018,chen2020,crinquand2020,2020ApJ...902...80K}. These PIC simulations typically neglect modeling the accretion disk and assume simple configurations, such as a monopolar magnetic field and an isotropic background photon field. However, the results of the PIC studies suggest that the VHE luminosity and spectrum depend crucially on the surrounding magnetic and radiation field structure. Therefore, it is necessary to consider them in the context of global GRMHD simulations in order to make meaningful predictions about the VHE $\gamma$-ray emission from M87.

Here we take a step towards bridging global calculations of accretion onto M87 with 
local kinetic models of gap acceleration. We carry out three-dimensional, global, radiation GRMHD simulations of M87 using the public \textsc{ebhlight} code \citep{ryan2015,ryan2019}. The frequency-dependent radiation field is evolved using a Monte Carlo method. We use the simulations to study the fluid and radiation field in the dense accretion flow (\autoref{sec:sims}) and evacuated funnel (\autoref{sec:funnel}) regions. We next study models of plasma production in the funnel, either through pair production by photons emitted in the accretion flow (\autoref{sec:drizzle}) or by high energy $\gamma$-ray photons emitted by particles accelerated in a gap (\autoref{sec:lgap}). Our results have implications for both PIC simulations of kinetic processes in the magnetosphere and the feasibility of near horizon particle acceleration from M87 (\autoref{sec:discussion}). 

\begin{figure*}
    \begin{tabular}{cc}
    \includegraphics[width=\columnwidth]{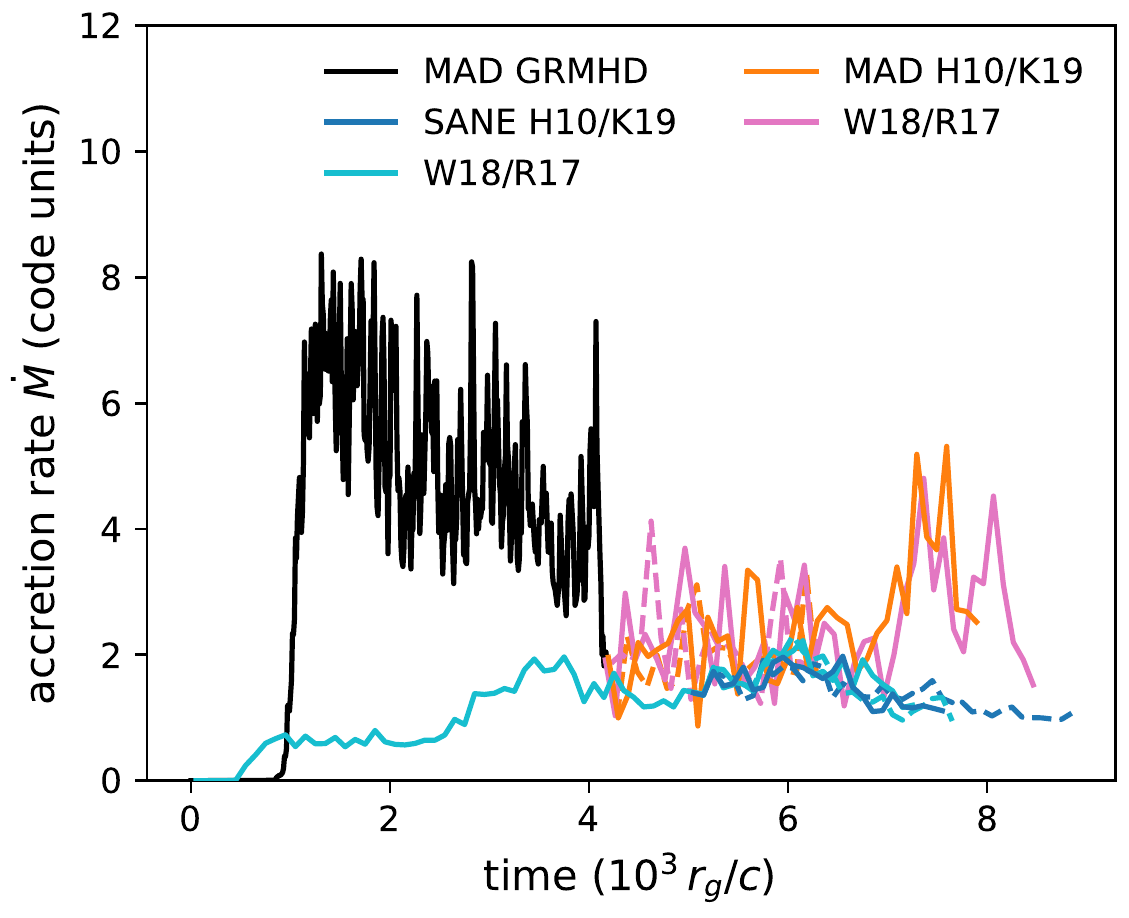} &
    \includegraphics[width=\columnwidth]{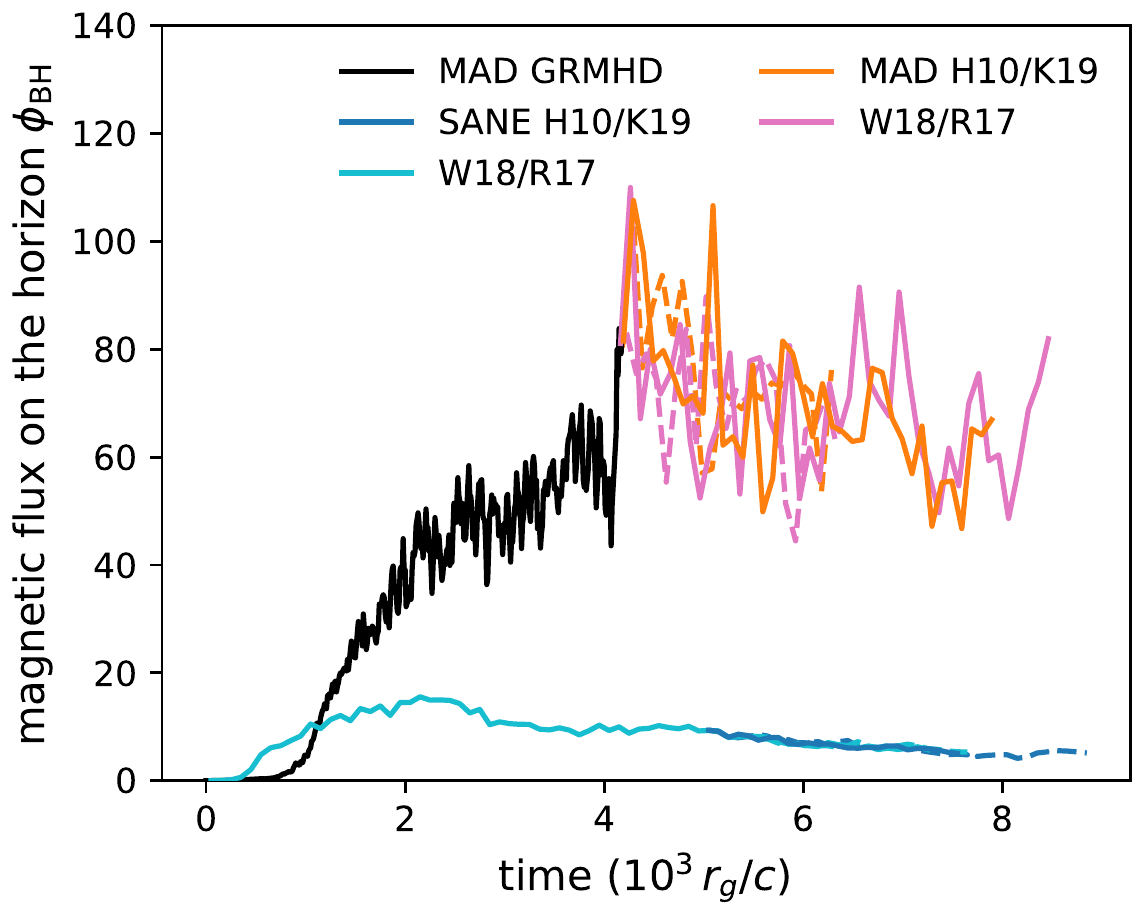}
    \end{tabular}
    \caption{Time series of mass accretion rate and dimensionless magnetic flux from the simulations as measured at the event horizon. The MAD radiation GRMHD simulations were all restarted from an initial non-radiative GRMHD calculation at a time $t \simeq 4\times10^3 \, r_g/c$ (black lines).}
    \label{fig:sim_plots_mdotphibh_t}
\end{figure*}

\begin{table*}
    \centering
    \caption{Properties of the \textsc{ebhlight} simulations in this work averaged over the final $2000 \, r_g/c$ of each simulation ($t \simeq 6-8 \times 10^3 \, r_g/c)$. The dimensionless magnetic flux $\phi_{\rm BH}$ is measured at the event horizon, and the other quantities are averaged over $4-10 \, r_g$. Here, $\langle Q \rangle$ denotes the average MRI resolution quality factors, $\langle \beta \rangle$ and $\langle \beta_{\rm rad} \rangle$ are the plasma and radiation $\beta$, and $N_{\rm sph}$ is the number of superphoton packets.}
    \label{tab:sim_table}
    \begin{tabular}{llccccccccccc}
    \hline
magnetic field & electron model & $\phi_{\rm BH}$ & $\langle B_r^2 \rangle / \langle B_\phi^2 \rangle$ & $\langle Q_\theta \rangle$ & $\langle Q_\phi \rangle$ & $\langle \beta \rangle$ & $N_{\rm sph}$ ($10^7$) & $\langle Q_{\rm emit} \rangle$ & $\langle \beta_{\rm rad} \rangle$\\
\hline
SANE & H10 & 6.3 & 0.15 & 22.5 & 23.6 & 24.7 & 7.9 & 225.0 & 145.2\\
SANE & W18 & 6.6 & 0.15 & 25.2 & 26.2 & 19.1 & 4.2 & 347.5 & 33.0\\
SANE & R17 & 6.4 & 0.14 & 23.6 & 25.0 & 21.1 & 3.9 & 252.6 & 55.5\\
SANE & K19 & 5.1 & 0.14 & 18.9 & 20.2 & 30.9 & 3.7 & 318.8 & 773.0\\
MAD & H10 & 64.7 & 0.26 & 70.6 & 64.3 & 5.7 & 7.1 & 12.3 & 9.4\\
MAD & W18 & 67.6 & 0.23 & 61.0 & 56.1 & 7.7 & 6.5 & 110.5 & 18.5\\
MAD & R17 & 71.4 & 0.33 & 121.8 & 84.7 & 4.3 & 4.9 & 19.8 & 9.8\\
MAD & K19 & 73.1 & 0.26 & 96.6 & 79.8 & 4.4 & 5.8 & 15.7 & 12.1\\
\hline
    \end{tabular}
\end{table*}

\begin{figure*}
    \begin{tabular}{c}
    \includegraphics[width=\textwidth]{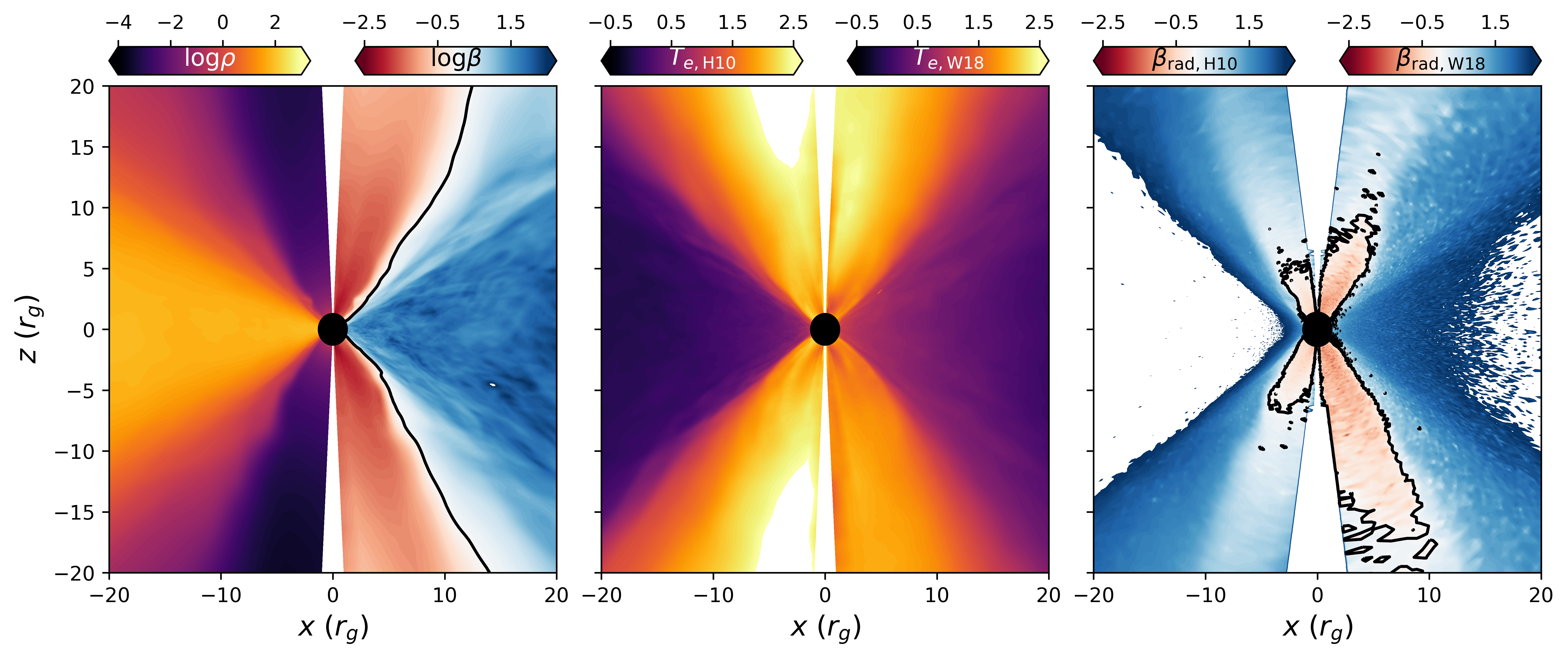}\\
        \includegraphics[width=\textwidth]{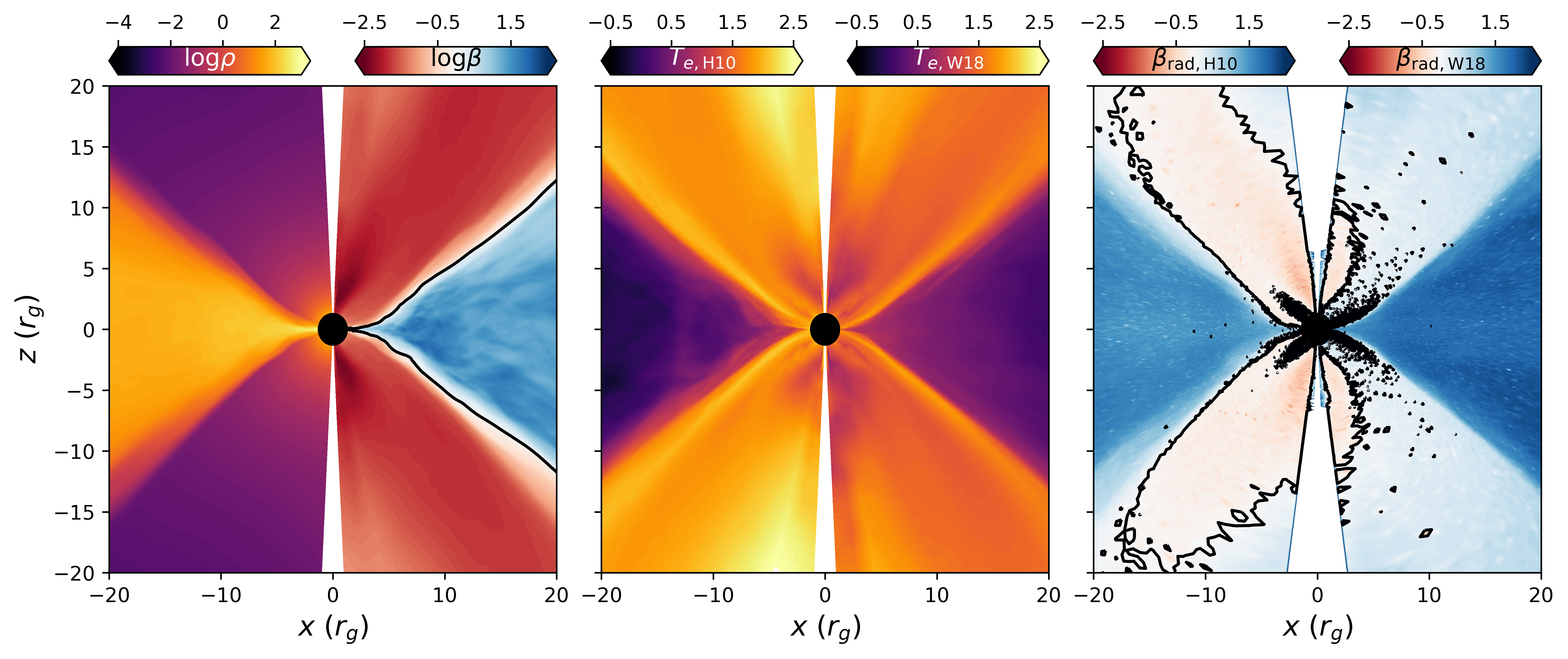}
    \end{tabular}
    \caption{Azimuthally averaged mass density $\rho$ (in code units) and ratio of gas to magnetic pressure $\beta$ (left), electron temperatures $T_e$ for the H10 and W18 heating models (in units of $m_e c^2 / k$, middle), and gas to radiation pressure ratio $\beta_{\rm rad}$ (right) for SANE (top) and MAD (bottom) simulations. We show only the W18 maps of $\rho$ and $\beta$, since they are indistinguishable by eye from those of the H10 simulations. The black solid lines mark where $\beta = 1$.}
    \label{fig:sim_plots_rad2d}
\end{figure*}

\section{Simulations}
\label{sec:sims}

We carried out radiation GRMHD simulations using version 1.0 of the public code \textsc{ebhlight}\footnote{\url{https://github.com/AFD-Illinois/ebhlight}} \citep{ryan2015,ryan2017}. \textsc{ebhlight} uses the \textsc{HARM} scheme for conservative ideal GRMHD \citep{gammie2003,noble2006} and implements a Monte Carlo treatment of the anisotropic, frequency-dependent radiation field. The electron entropy is evolved separately, with a heating rate calculated as a fraction of the dissipated grid-scale energy according to a sub-grid prescription \citep{ressler2015}. Synchrotron emission and self-absorption and Compton scattering are included self-consistently.\footnote{We did not include bremsstrahlung, which is sub-dominant in terms of cooling but could contribute to the high energy component of the SED \citep{yarza2020}.} The integrations are done explicitly with a time step limited by the light-crossing time of the smallest zone. We find that the fiducial simulations presented here are $\simeq 4-5\times$ more computationally expensive than non-radiative GRMHD simulations performed at the same grid resolution. 

\subsection{Simulation setup}

The simulations were initialized from a Fishbone--Moncrief torus with an inner radius of $r_{\rm in} = 12 \, r_g$, a pressure maximum radius of $r_{\rm max} = 25 \, r_g$, and a dimensionless black hole spin parameter $a \equiv Jc / GM^2 = 0.9375$, where $r_g = GM/c^2$, $J$ is the angular momentum of the hole, and $M_{\rm BH} = 6.5 \times 10^9 M_\odot$ \citep{EHTPaperVI,gebhardt2011}. The grid resolution was $320 \times 256 \times 160$, and simulations were carried out in modified Kerr-Schild coordinates\citep[]{porth2019}, which concentrate $\theta$ resolution towards the midplane at small radius and allow a larger timestep. The torus is seeded with a magnetic field chosen to accumulate weak (SANE) or strong (dynamically important, MAD) magnetic fields on the central black hole. We used four separate schemes for electron heating, based on the results of kinetic calculations. We considered turbulent heating as realized in gyrokinetics calculations \citep[H10 and K19,][]{howes2010,kawazura2019} and heating by magnetic reconnection as realized in particle-in-cell calculations with a modest guide field \citep[R17 and W18,][]{rowan2017,werner2018}. The gyrokinetic turbulence prescriptions introduce a strong magnetization dependence into the electron heating fraction (electrons receive a greater fraction of heating in more strongly magnetized regions), while the reconnection prescriptions are more uniform \citep[see e.g.,][]{chael2019}. In the non-radiative case, multiple electron species can be evolved in parallel to explore different heating models during a single simulation \citep[e.g.,][]{ressler2017}. Here, a new simulation is required for each sub-grid prescription since the radiation field is coupled to the MHD and depends on the electron temperature. A ceiling was imposed on the magnetization parameter, $\sigma = b^2 / \rho \le 50$, to allow for stable evolution of the magnetically dominated polar regions. This procedure artificially injects mass at high temperature. To avoid generating luminosity from this injected material, regions where $\sigma > 1$ or the dimensionless electron temperature $k T_e / m_e c^2 > 10^3$ do not contribute to the emission, absorption, or scattering.

We first ran an initial, non-radiative MAD GRMHD simulation to $t = 4.2 \times 10^3 \, r_g/c$ using the W18 heating prescription. The radiation was then turned on, and we ran four simulations using each of the H10/K19/W18/R17 electron heating prescriptions. The simulation mass unit was chosen to approximately match the observed $\simeq 1$ Jy $230$ GHz flux density of M87. This choice determines the time-averaged mass accretion rate onto the black hole. After turning on radiation, the system re-equilibrates after a dynamical time of $\lesssim 10^2 \, r_g/c$. For the SANE simulations, we started with a radiative GRMHD simulation also using the W18 prescription. We then restarted from that solution at $t = 5 \times10^3 \, r_g/c$, again running four simulations using each of the four electron heating prescriptions. The procedure used for the MAD case is more efficient, since at early times a longer timestep can be taken in the absence of radiation and the execution time per timestep is shorter. All simulations were run for a total duration of $\simeq 8 \times 10^3 \, r_g/c$. 

\autoref{fig:sim_plots_mdotphibh_t} shows the time evolution of the (positive) mass accretion rate $\dot{M}$ and the dimensionless magnetic flux on the horizon $\phi = \sqrt{4\pi} \Phi / \sqrt{\dot{M}}$, both in code units with $GM = c = 1$. In the MAD case, the magnetic flux ramps up and saturates at a value of $\phi \simeq 60-80$, consistent with past results \citep{tchekhovskoy2011,mckinney2012}. The accretion rate quickly reaches a maximum, then slowly decreases. We note that the large drop around $\simeq 4 \times 10^3 \, r_g/c$ occurs before radiation is turned on, and we occasionally see similar features in other, non-radiative MAD GRMHD simulations \citep[e.g.,][]{dexter2020harmpi}. In the SANE case, the accretion rate reaches a maximum more slowly since the magnetorotational instability \citep[MRI,][]{mri} is not well resolved in the outer torus where most of the mass is contained \citep[e.g.,][]{liska2018}. The magnetic flux saturates at a lower value of $\phi \simeq 5$. We define the equilibrium radius as the outermost location where the elapsed time is longer than the local inflow time, $t_{\rm inflow} = r/|v^r|$ \citep{narayan2012}. According to that criterion, inflow equilibrium is established out to $r_{\rm eq} \simeq 15 \, r_g$, $13 \, r_g$ for the MAD and SANE simulations by $7\times 10^{3} \, r_g/c$.

The radiation field in \textsc{ebhlight} comprises a large number of Monte Carlo samples (superphotons), each with a particular position and wavevector and corresponding to a large number of individual photons. 
One quality factor for the radiation field is the number of superphotons emitted locally in one cooling time due to synchrotron radiation \citep{miller2019}, $Q_{\rm emit} = t_{\rm cool} \, N_{\rm emit} / \Delta t$. Here, the cooling time $t_{\rm cool} = u_g / \Lambda$ is the ratio of the energy density of the gas to the cooling rate, and the average rate of superphoton emission
is $N_{\rm emit} / \Delta t$. A minimal value of $Q_{\rm emit} \gtrsim 1$ is needed to avoid supercooling events, during which a simulation zone cools too quickly and releases a significant fraction of its internal energy in a single simulation timestep. In all of our simulations, we find density-weighted shell-averaged values $\langle Q_{\rm emit} \rangle \simeq 10-200$ near the black hole (\autoref{tab:sim_table}). An analogous quality factor can be defined for scattering events (inverse Compton cooling). This factor decreases rapidly with decreasing radius, becoming $< 1$ very close to the event horizon. We study the properties of both quality factors in Appendix \ref{app:nph_converge} as a function of the number of superphotons used in the calculation. The values found here appear satisfactory to provide a converged description of the time- and shell-averaged radiation field. In appendix \ref{app:coolcompare}, we show that radiative cooling reduces the average near-horizon electron temperature by factors of $\simeq 2-5$.

Azimuthally averaged snapshot 2D maps of selected fluid and radiation quantities from the H10/W18 electron heating models are shown in \autoref{fig:sim_plots_rad2d}. The H10 prescription results in a colder accretion flow and hotter funnel. The electron temperatures of our MAD models are somewhat higher than their SANE counterparts. Radiation is more important than gas pressure over a large swath of the funnel for the MAD H10 and SANE W18 cases, and the gas pressure is in general overestimated there due to the influence of numerical floors. Our SANE H10 simulation radiates a lower luminosity by a factor of $\simeq 2$. If we increased $\dot{M}$ to match the luminosity of the other simulations, the effect of radiative cooling would be stronger than found here.

\begin{figure*}
    \begin{tabular}{cc}
    \includegraphics[width=0.5\textwidth]{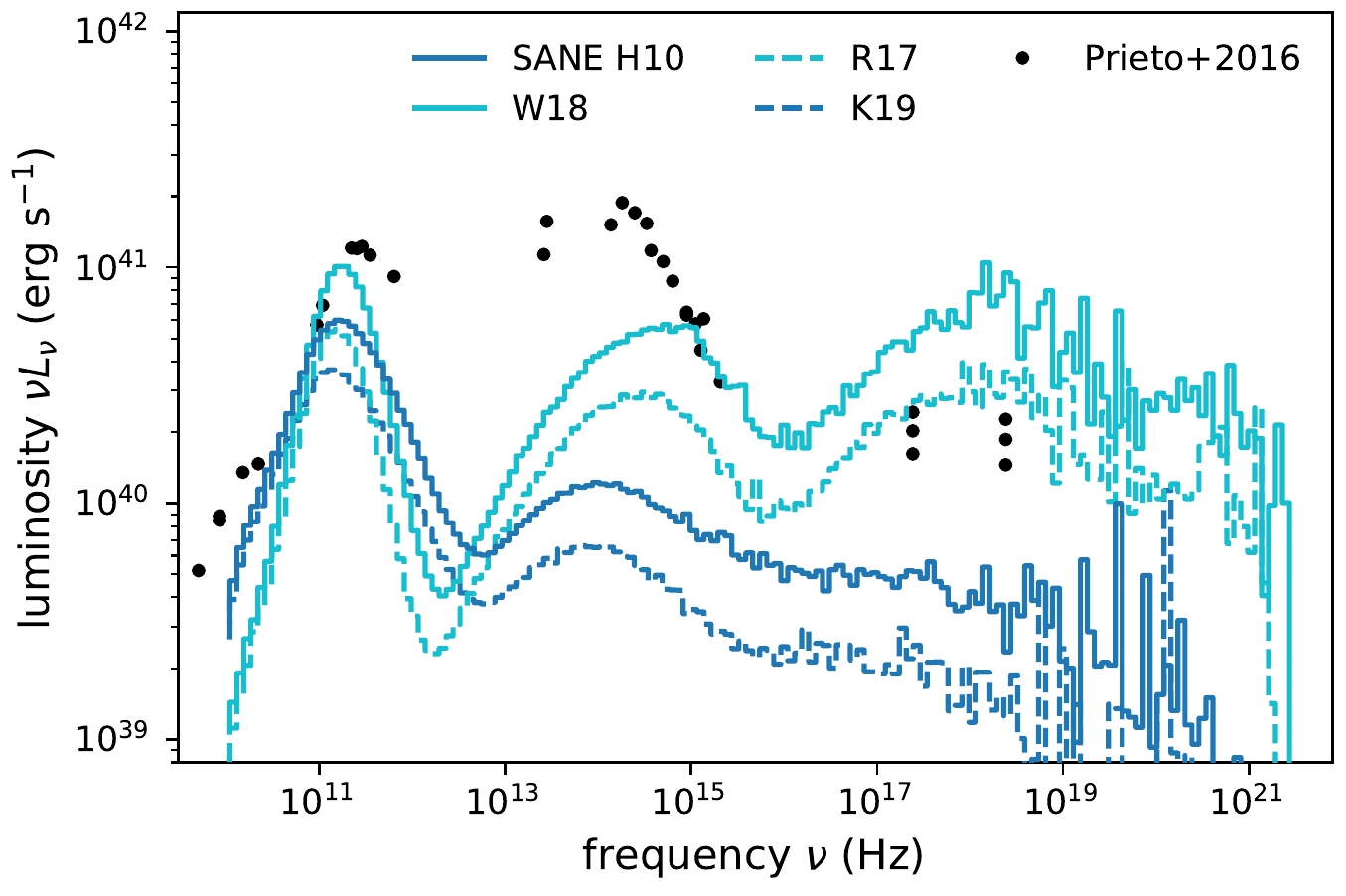} &
        \includegraphics[width=0.5\textwidth]{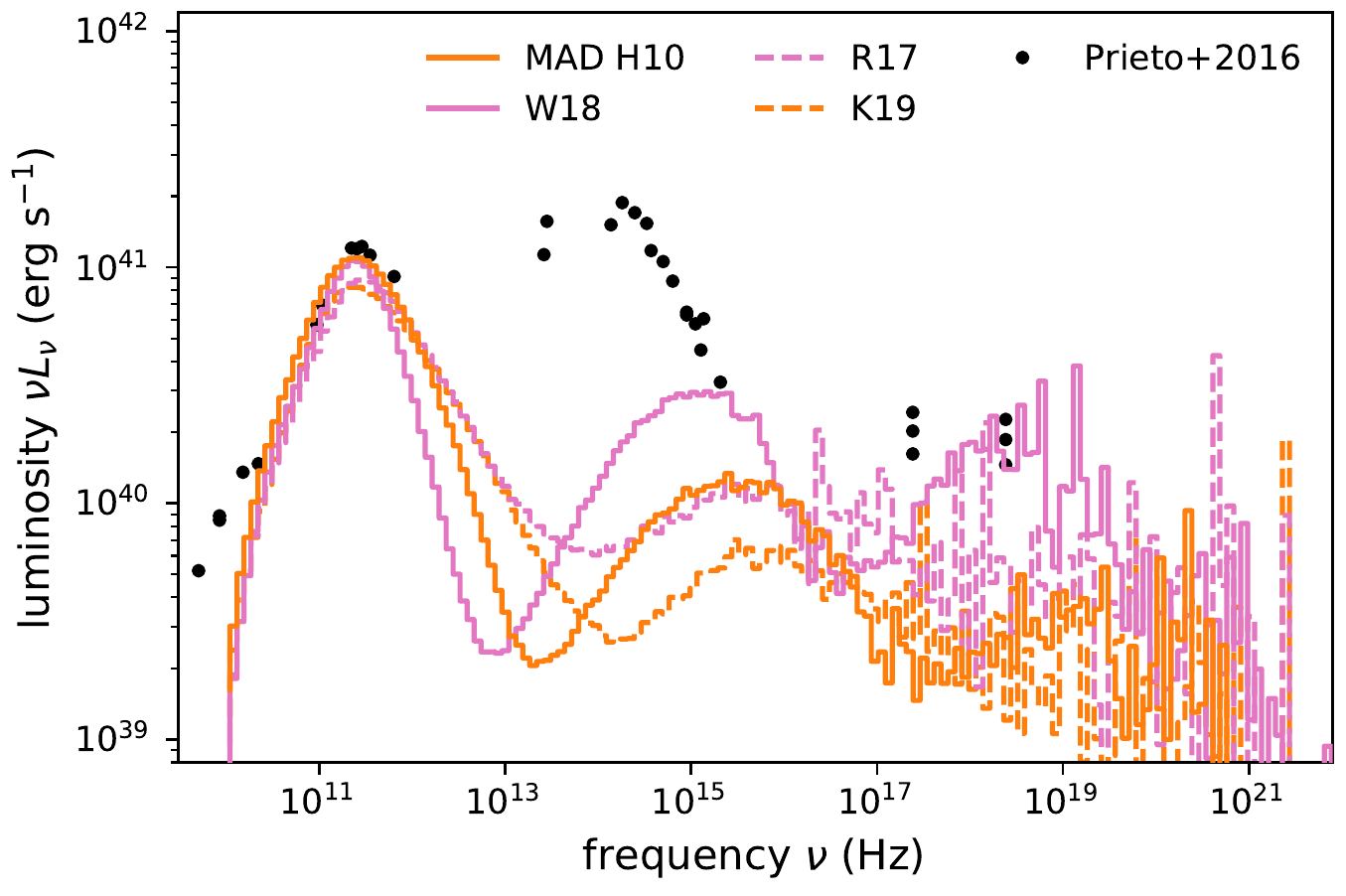}
    \end{tabular}
    \caption{Spectra averaged over the last $500 \, r_g/c$ from our SANE (left) and MAD (right) models compared to M87 data (\citealt{prieto2016} and references therein). The MAD R17 model also shows significant spectral variability, which causes a temporary increase in the infrared luminosity by a factor of several above the quiescent average shown here.}
    \label{fig:sim_plots_spectra}
\end{figure*}

We calculate observed spectra from our simulations by collecting superphoton packets at a radius $r = 40 \, r_g$. \autoref{fig:sim_plots_spectra} shows spectra for a low inclination observer. The spectra are binned over the final $500 \, r_g/c$ of simulation time, over all azimuthal angles, and over low inclination angles relative to the black hole spin axis of $0-25^\circ$
to be consistent with the inferred jet viewing geometry of M87 \citep[e.g.,][]{biretta1999}. Synchrotron radiation produces an initial peak near THz frequencies, with a broader peak in models with a wider range of temperatures for the radiating electrons (SANE H10/K19 and MADs). The spacing of the Compton bumps is set by the typical electron temperature in the accretion flow, where the scattering optical depth is largest. The relatively hot and dense accretion flow in SANE W18/R17 models produces a large X-ray luminosity from inverse Compton scattering and exceeds the total observed X-ray luminosity of M87 \citep[data points compiled by][see references therein]{prieto2016}. With a purely thermal distribution of electrons, our models can match the observed submm and potentially X-ray luminosity. They underproduce the infrared emission, which could arise from synchrotron jet \citep{perlman2001} or thermal accretion disk \citep{prieto2016} emission. In the case of synchrotron radiation, a non-thermal tail to the distribution function is likely responsible \citep{davelaar2019}. Mass accretion rate values, bolometric luminosities, jet powers, and their efficiencies are listed in \autoref{tab:rad_table}. Our high spin MAD models have radiative efficiencies of $\simeq 10\%$ and jet efficiencies $> 100\%$. The high jet efficiencies are consistent with past work \citep[e.g.,][]{tchekhovskoy2011} and demonstrate spin energy extraction from the black hole.

\begin{table*}
    \centering
    \caption{Average properties of the \textsc{ebhlight} simulations presented here averaged over the final $2000 \, r_g/c$ of each simulation. The accretion rate $\dot{M}$ is measured at the event horizon. The bolometric luminosity is measured at $40 \, r_g$, and the jet power $L_{\rm jet}$ is measured at $100 \, r_g$. $L_{\rm scat}/L_{\rm emit}$ is the ratio of luminosities produced by scattering to those produced by emission. Luminosities $L_{\rm 42}$ are reported in units of $10^{42} \, \rm erg \, \rm s^{-1}$, and $\dot{M}_4$ is the mass accretion rate in units of $10^{-4} \, M_\odot \, \rm yr^{-1}$.}
    \label{tab:rad_table}
    \begin{tabular}{llcccccc}
    \hline
magnetic field & electron model & $\dot{M}_{-4}$ & $L_{\rm scat}/L_{\rm emit}$ & $L_{\rm bol, 42}$ & $L_{\rm jet, 42}$ & $L_{\rm bol}/\dot{M}c^2$ & $L_{\rm jet}/\dot{M}c^2$\\
\hline
SANE & H10 & 7.2 & 0.8 & 0.4 & 1.0 & 0.009 & 0.024\\
SANE & W18 & 8.1 & 4.3 & 1.2 & 0.9 & 0.026 & 0.019\\
SANE & R17 & 6.8 & 3.8 & 0.9 & 1.5 & 0.023 & 0.040\\
SANE & K19 & 5.7 & 0.6 & 0.2 & 0.9 & 0.006 & 0.027\\
MAD & H10 & 1.6 & 0.4 & 1.0 & 17.1 & 0.111 & 1.836\\
MAD & W18 & 3.5 & 1.2 & 1.5 & 40.2 & 0.075 & 2.020\\
MAD & R17 & 3.0 & 0.9 & 1.6 & 38.9 & 0.095 & 2.268\\
MAD & K19 & 1.2 & 0.3 & 0.6 & 17.0 & 0.094 & 2.461\\
\hline
    \end{tabular}
\end{table*}

\begin{figure*}
    \begin{tabular}{cc}
    \includegraphics[width=0.7\textwidth]{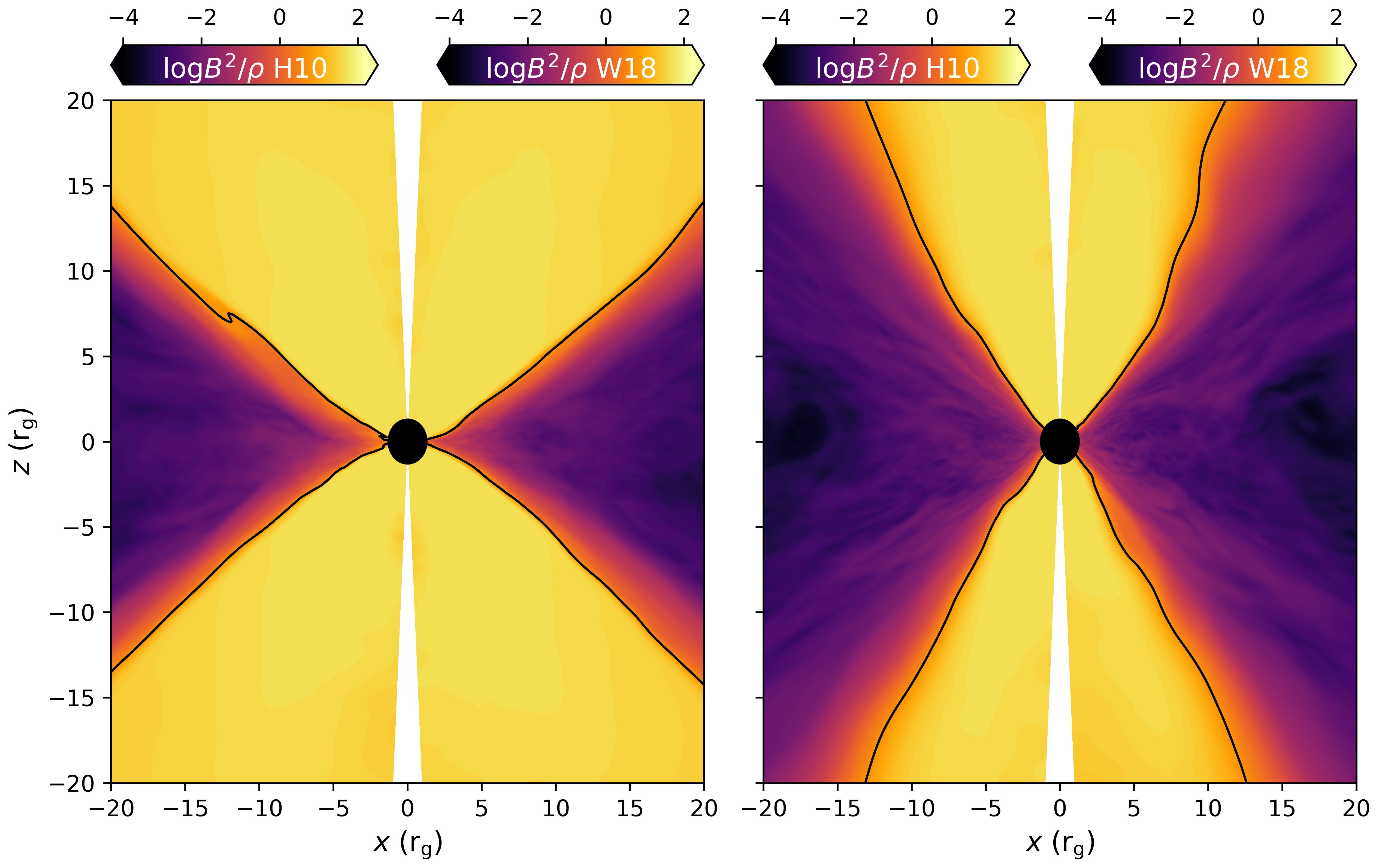}
    \end{tabular}
    \caption{Azimuthally averaged magnetization (ratio of magnetic to rest mass energy density) for sample MAD (left) and SANE (right) simulations. The black solid line marks our adopted definition of the jet boundary ($\sigma \geqslant 5$).}
    \label{fig:funnel_contour}
\end{figure*}

\begin{table*}
    \centering
    \caption{Photon emission region location and size in radius and polar angle for sample MAD and SANE simulations. The SANE case is split into contributions from above (North) and below (South) the equatorial plane.}
    \label{tab:cooling_table}
    \begin{tabular}{llcccc}
    \hline
magnetic field & electron model & $\rm r_c$ ($r_g$) & $\rm \Delta r_c$ ($r_g$) &  $\rm \theta_c$ & $\rm \Delta \theta_c$ \\
\hline
SANE (North) & H10 & 1.7 & 0.39 & 0.41$\rm \pi$ & 0.05$\rm \pi$\\
SANE (South) & H10 & 1.7 & 0.40 & 0.57$\rm \pi$ & 0.05$\rm \pi$\\
MAD & W18 & 1.7 & 0.49 & 0.51$\rm \pi$ & 0.03$\rm \pi$\\
\hline
    \end{tabular}
\end{table*}

\section{Funnel magnetic and radiation field properties}
\label{sec:funnel}

\begin{figure*}
    \begin{tabular}{cc}
    \includegraphics[width=0.7\textwidth]{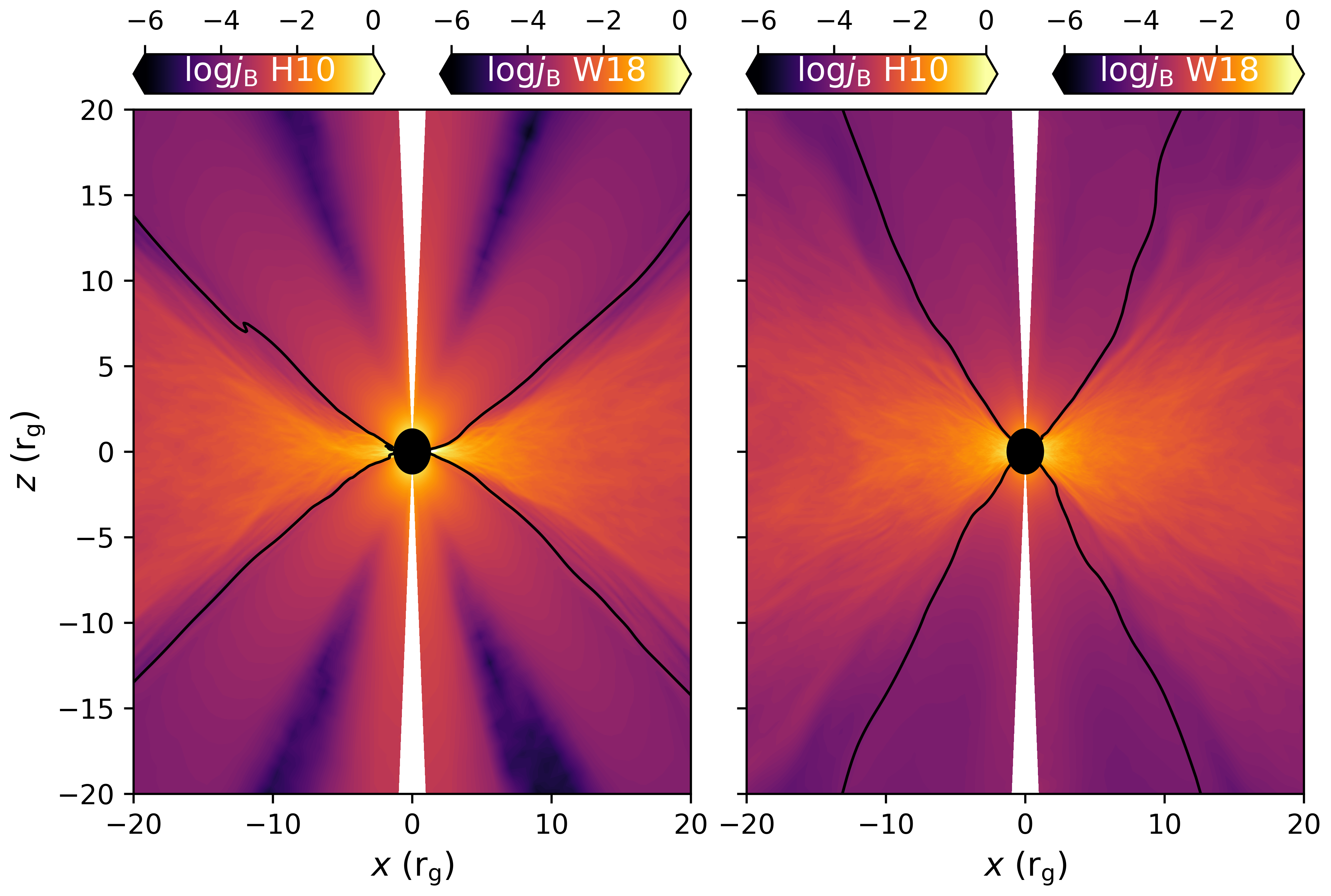}
    \end{tabular}
    \caption{Log contour plots of background current density for MAD simulations (left) and SANE simulations (right) with H10 or W18 electron heating schemes. The black solid line marks the jet boundary defined with $\sigma \geqslant 5$}
    \label{fig:jB_trend}
\end{figure*}

\begin{figure*}
    \begin{tabular}{cc}
    \includegraphics[width=\columnwidth]{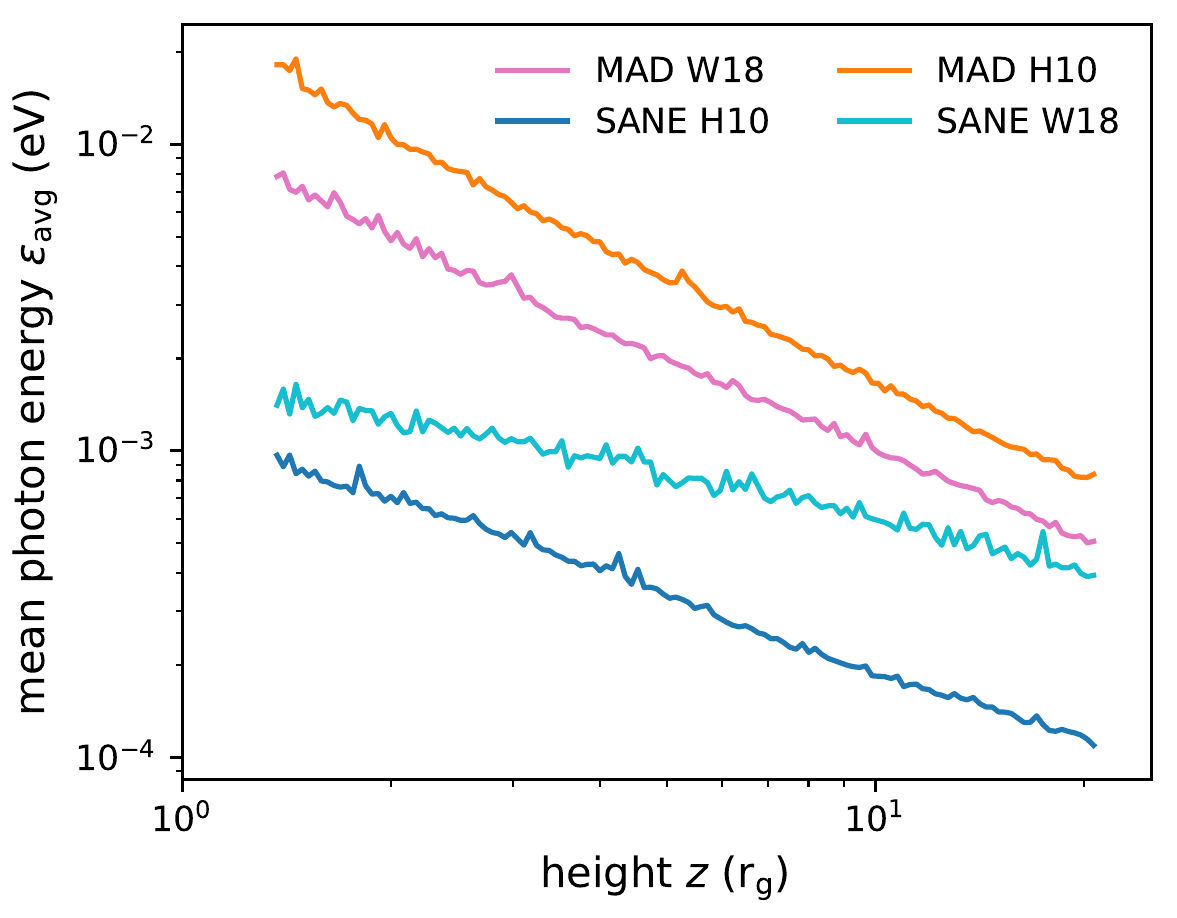}
    \includegraphics[width=\columnwidth]{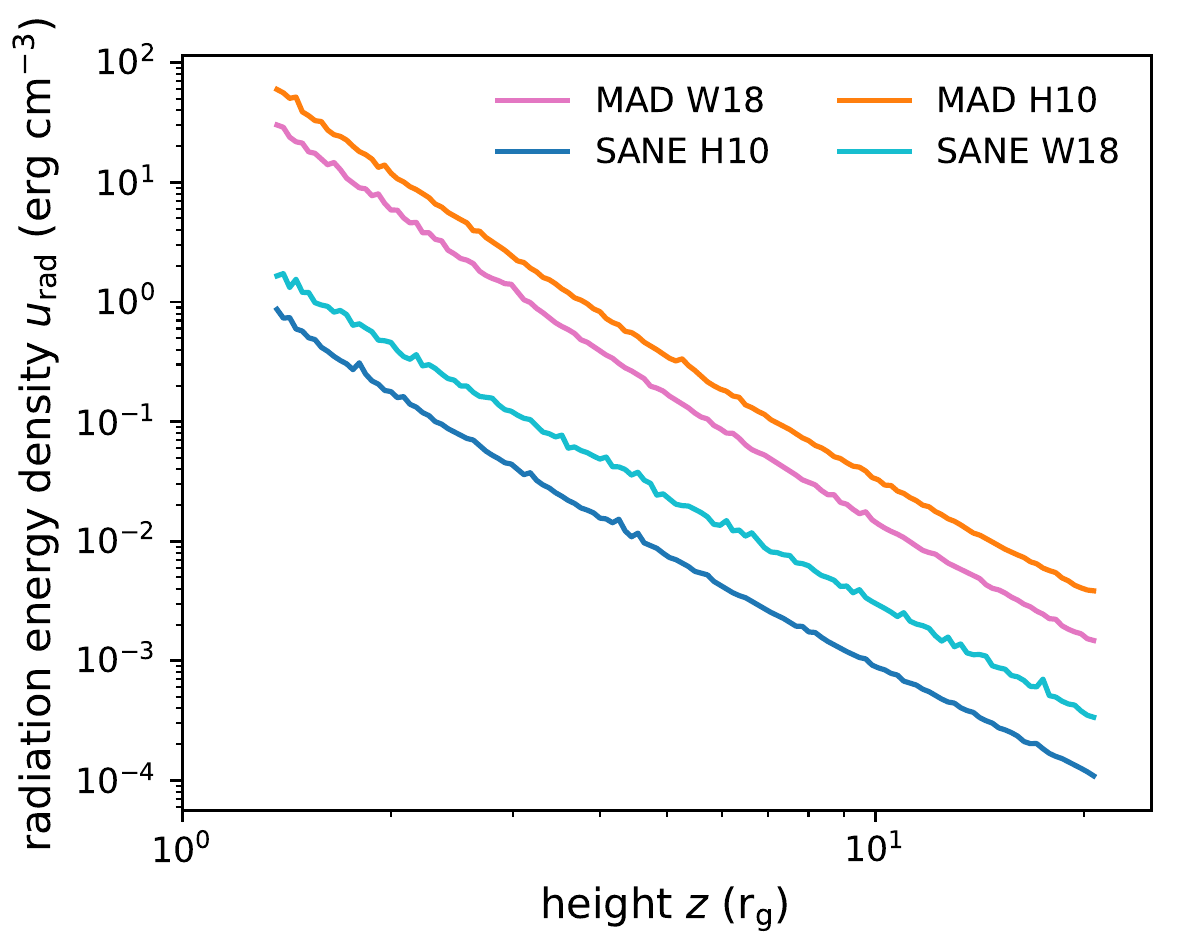}
    \end{tabular}
    \caption{Average photon energy ($\epsilon_{avg} = u_{\rm rad}/n_s$, left) and radiation energy density ($u_{\rm rad}$, right) plotted against height for MAD simulation with W18 (pink) or H10 (orange) electron heating schemes and SANE simulation with W18 (cyan) or H10 (blue) electron heating schemes.} 
    \label{fig:nph_ur}
\end{figure*}

In order to study the local particle acceleration inside the jet funnel, we need to understand properties of the radiation field and the magnetic field within the jet. We begin the analysis by defining a general jet boundary in our simulation results using an azimuthally averaged magnetization parameter $\sigma = \left<b^2\right>/\left<\rho\right>$, defined with the magnetic field $b$ and matter density $\rho$. We define our jet to be the magnetically dominated region where $\sigma \geqslant \xi$. The choice $\xi = 1$ of \citet{mckinney2012} includes the turbulent accretion disc of our MAD simulations. Since it seems unlikely that an empty, unscreened gap could form there, we instead chose $\xi = 5$ (\autoref{fig:funnel_contour}). By our definition SANE simulations have a narrower funnel than MAD simulations. Because of the relatively mild differences produced by these electron heating schemes, we will use H10 and W18 to study the differences in the remainder of this section.

\subsection{Radiation energy density and magnetic field structure}

The radiation physics sector of the calculation is not scale free, so the length (and time) and density invariance of metric and GRMHD evolution equations must be broken. In our simulation, the relevant length and time scales are both set by the mass of the central black hole: $r_g$ and $r_g/c$. We set the final (density) scale of each simulation by simultaneously rescaling the rest mass density, internal energy, and magnetic field strength until the simulations produce the observed $1.3$mm flux density of M87 ($\simeq 1$ Jy for $D = 16.8$ Mpc, \citealt{tully1988}). We compute $\langle x \rangle_f \, (z)$, the height-dependent, funnel-averaged profile of a variable $x$, by calculating its shell-average:
\begin{equation}
 \langle x \rangle_f = \dfrac{\sum^{2\pi}_{j=0}\sum^{\theta_{k2}}_{k=\theta_{k1}} x_{jk} \sqrt{-g_{jk}}} {\sum^{2\pi}_{j=0}\sum^{\theta_{k2}}_{k=\theta_{k1}} \sqrt{-g_{jk}}},
\end{equation}
\noindent where $\sqrt{-g}$ is the metric determinant.

We extract the photon number density $n_s$, the radiation energy density $u_{\rm rad}$, the background current density squared $j^2 = j^\mu j_\mu$ ($j^\mu = {F^{\mu \nu}}_{;\nu}$ and $F^{\mu \nu}$ is the electromagnetic field tensor), and the magnetic field strength squared $b^2 = b^\mu b_\mu$ ($b^\mu$ is the magnetic field 4-vector) from the simulation data. These quantities are converted to cgs units and further multiplied by a factor of $\sqrt{4\pi}$ to convert from the Heaviside--Lorentz to Gaussian field convention.

\autoref{fig:jB_trend} shows azimuthally averaged maps of the magnitude of the current density. 
The choice of electron heating scheme has little effect on the background current density because the radiation field is not strong enough to influence it. The current density is a factor of several higher in the MAD simulations than in the SANE ones.

\autoref{fig:nph_ur} shows vertical profiles of the radiation energy density and average photon energy. Both decay as power laws with height. At small radii, the radiation energy density is higher than simple estimates predicted by the observed luminosity assuming isotropy, e.g., $u_{\rm rad}$ increases more quickly towards the black hole than $\propto z^{-2}$. This is likely due to relativistic effects, e.g., the gravitational redshift. At $\sim 3 r_g$, the photon number density $n_s \simeq 10^{14-15}$ cm$^{-3}$ and $u_{\rm rad} \simeq 0.1-10 \, \rm erg \, \rm cm^{-3}$. The high end of the ranges corresponds to MAD simulations.

\begin{figure*}
    \begin{tabular}{cc}
    \includegraphics[width=0.8\columnwidth]{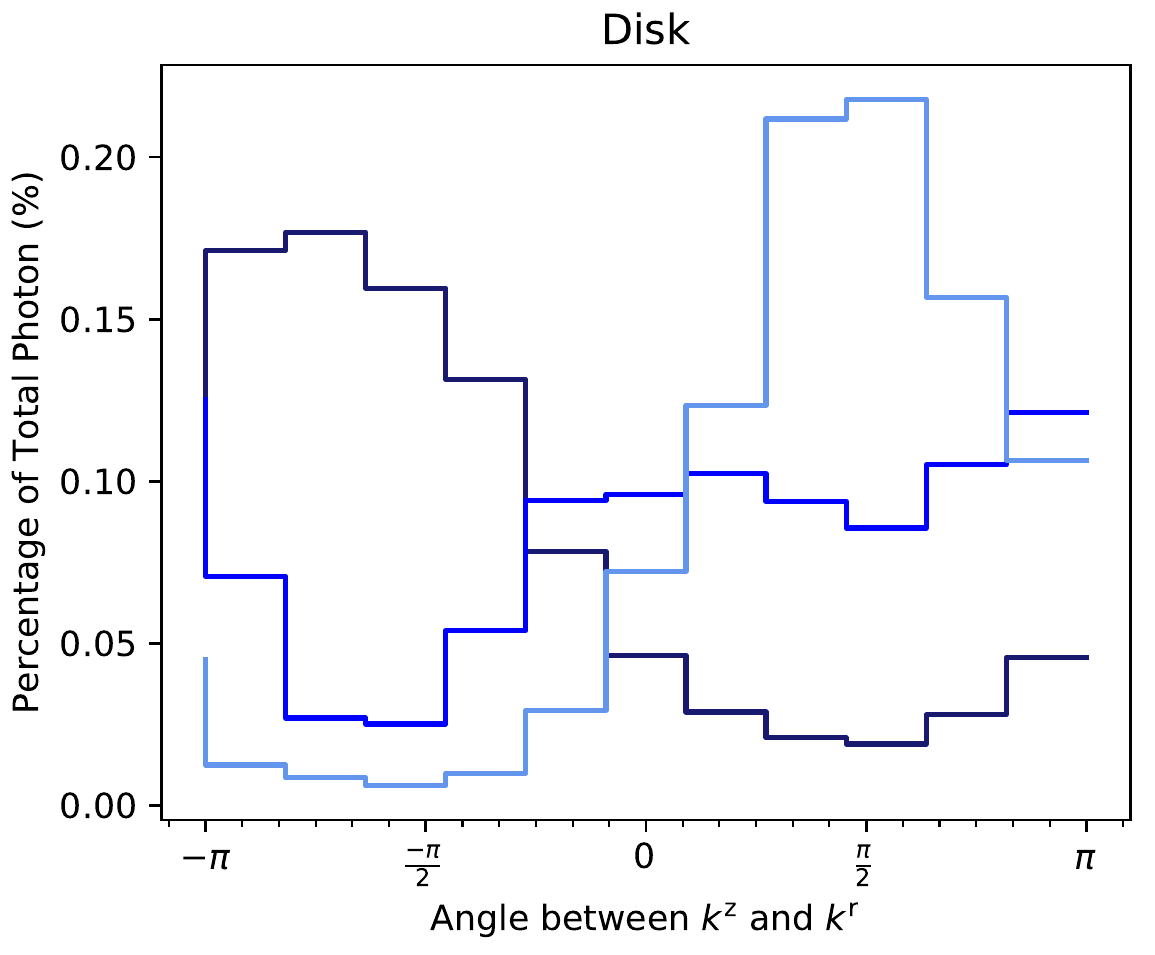}&
    \includegraphics[width=0.8\columnwidth]{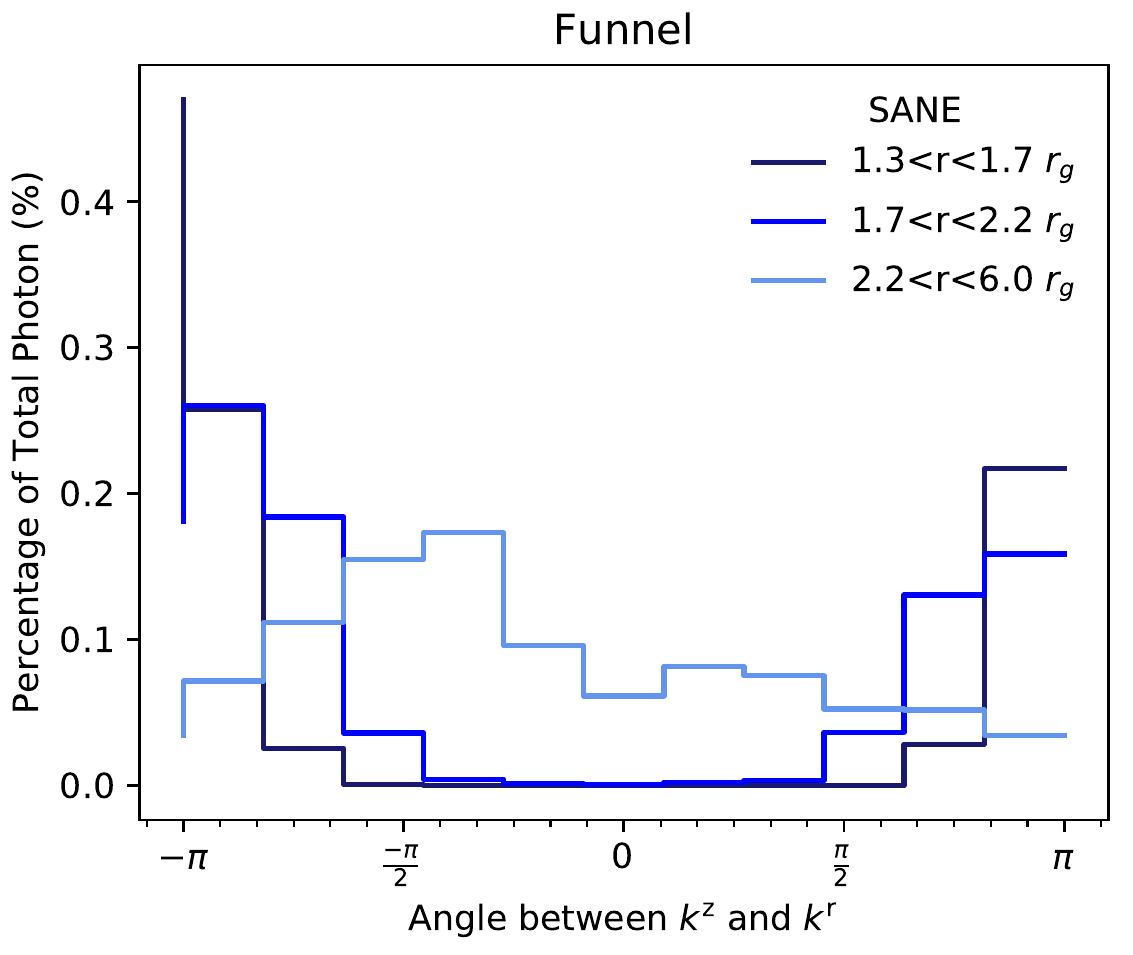}\\
    \includegraphics[width=0.8\columnwidth]{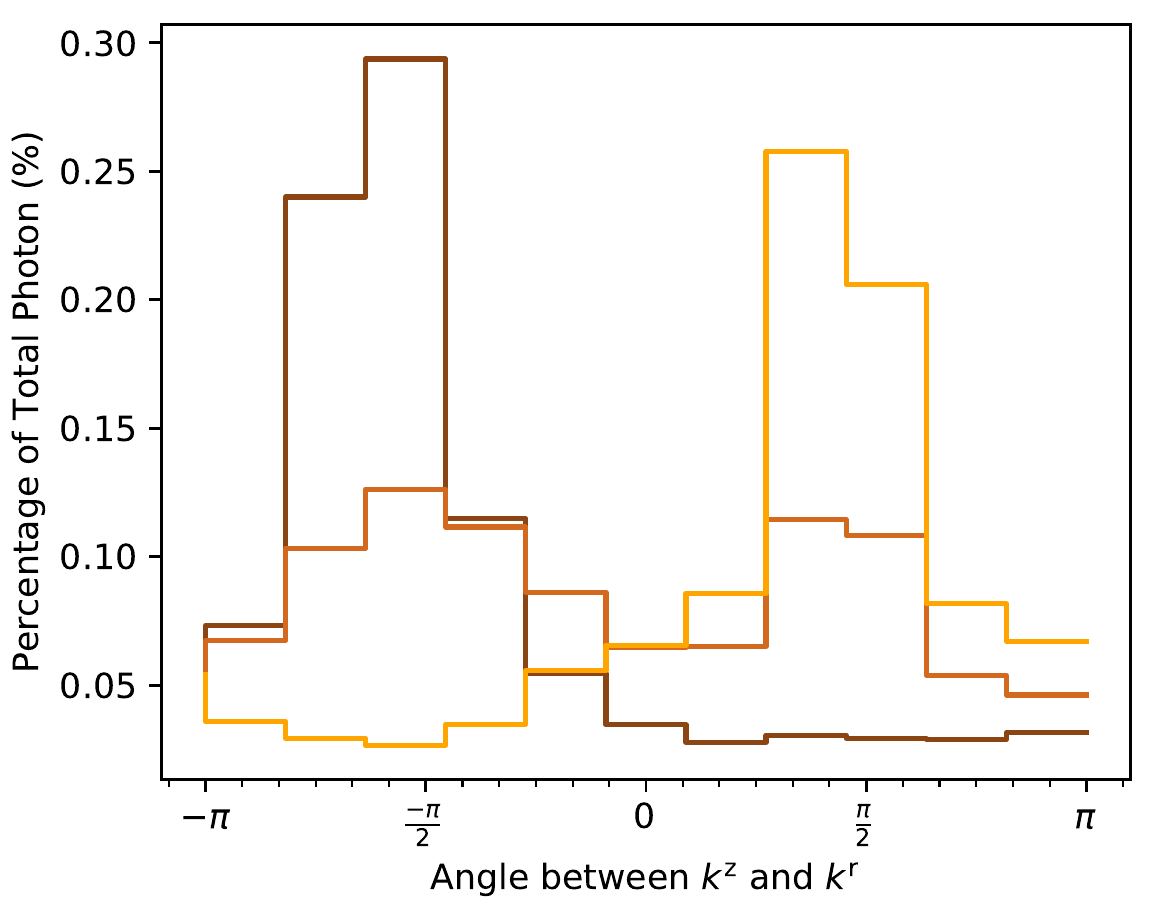}&
    \includegraphics[width=0.8\columnwidth]{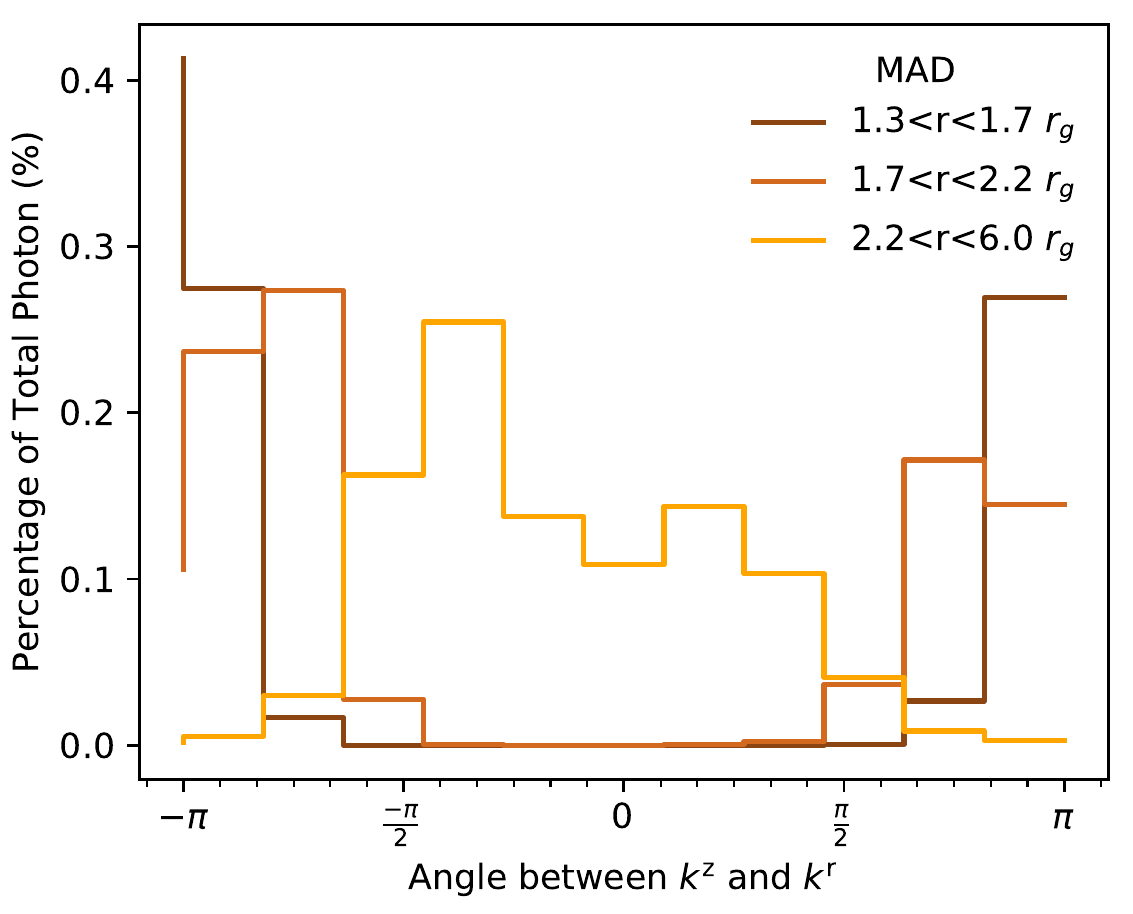}
    \end{tabular}
    \caption{Relative photon direction in height and cylindrical radius measured as $\alpha = {\rm sign}(k^{z}) \times {\rm cos^{-1}} (k^{z}/\sqrt{{k^{R}}^2+{k^{z}}^2)})$ in the funnel region and as $\alpha = {\rm sign}(k^{R}) \times {\rm cos^{-1}} (k^{z}/\sqrt{{k^{R}}^2+{k^{z}}^2)})$ in the equatorial plane, where $k^{i}$ are measured in the orthonormal ZAMO frame. The distributions are shown for SANE H10 (top) and MAD W18 (bottom) simulations in equatorial plane (left) and funnel region (right). Lighter colors represents greater radial distance of photon packets from the black hole. In general, the radiation field is anisotropic and varies from predominantly inwards to predominantly outwards trajectories between $1.4 \lesssim r \lesssim 6 \, r_g$.}
    \label{fig:photon_dist}
\end{figure*}

\begin{figure*}
    \includegraphics[width=0.8\textwidth]{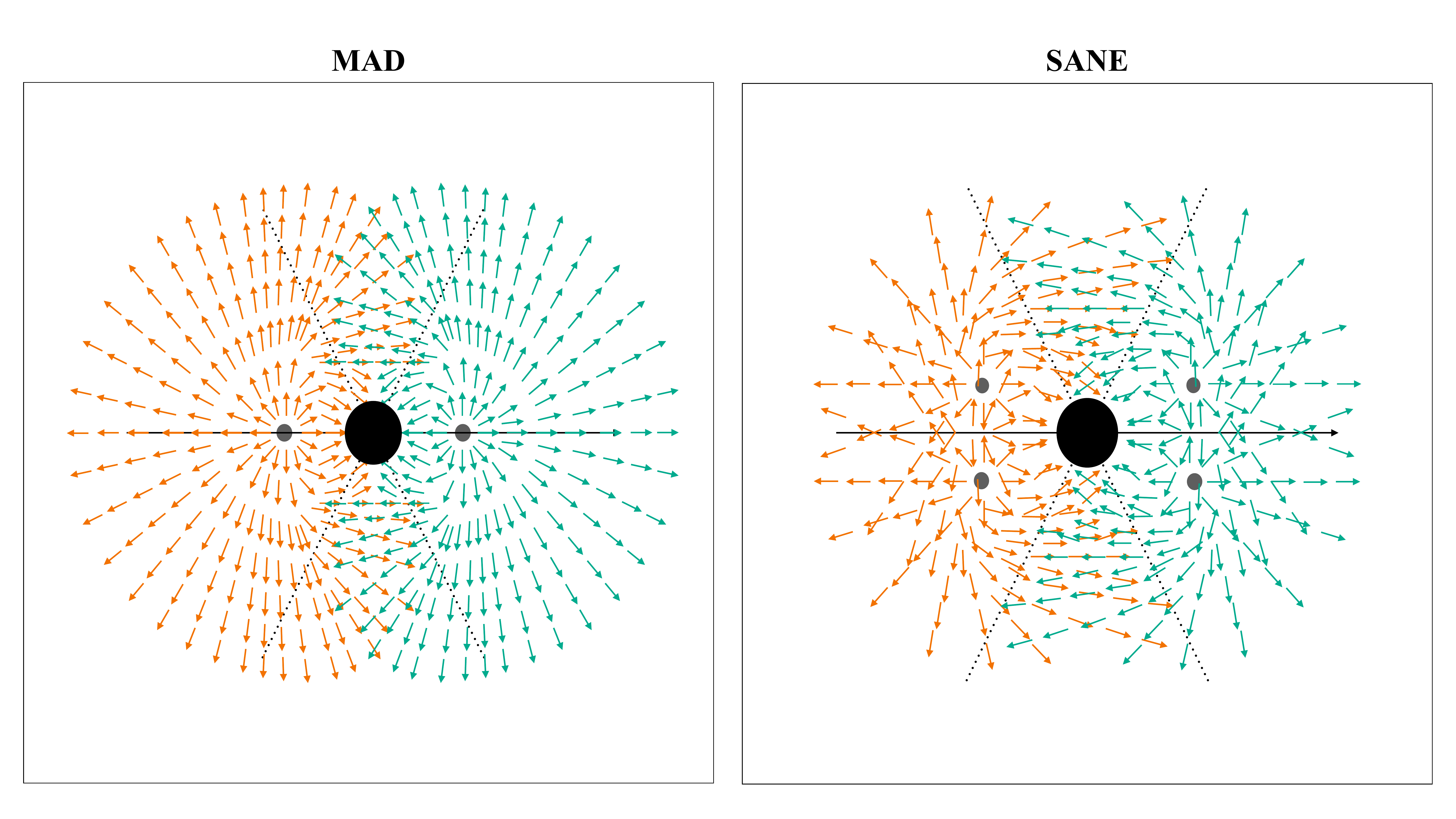}
    \caption{Schematic diagram showing the approximate photon distributions expected for emission from a ring at 1.7--2 $r_g$ in the equatorial plane (left) and for two similar rings elevated some height and symmetric about the equatorial plane. These toy geometries are motivated by the measured emission region properties for MAD (left) and SANE (right) simulations listed in \autoref{tab:cooling_table}, and help explain the anisotropic photon distributions shown in \autoref{fig:photon_dist}.}
    \label{fig:photon_dist_vector}
\end{figure*}

\subsection{Photon distribution}

We also measured the photon distribution in the funnel region and around the black hole. For a single time slice in each model, we output the coordinate positions $x^\mu$ and wave four-vectors $k^\mu$ of all superphoton packets in the simulation. We then transformed the four-vector components from the code coordinate basis to the orthonormal, zero angular momentum observer (ZAMO) \citep{bardeen1972} frame.

In addition to photon distributions, we calculate average cooling region locations and sizes in SANE and MAD simulations. In each simulation, we extract the volumetric radiative energy exchange rate due to emission $J_{\rm em}$ and find its mean and standard deviation in radius and polar angle, assuming axisymmetry, with:
\begin{equation}
\rm r_c =  {\sum_{r,\theta,\phi} J_{\rm em}\times r \over \sum_{r,\theta,\phi} J_{\rm em}},\\ \rm \Delta r_c = {\sum_{r,\theta,\phi} J_{\rm em}\times r^2 \over \sum_{r,\theta,\phi} J_{\rm em}},
\end{equation}
\noindent and similarly:
\begin{equation}
\rm \theta_c =  {\sum_{r,\theta,\phi} J_{\rm em}\times \theta \over \sum_{r,\theta,\phi} J_{\rm em}},\\ \rm \Delta \theta_c = {\sum_{r,\theta,\phi} J_{\rm em}\times \theta^2 \over \sum_{r,\theta,\phi} J_{\rm em}}.
\end{equation}

As shown in \autoref{tab:cooling_table}, the average photon emission radii for all models are very close to the event horizon. Photons in the MAD W18 simulation are emitted in a narrow polar angle range around the equatorial plane. In the SANE H10 simulation, we find that photons are emitted by two similar structures above and below the disk. This is due to the fact that the H10 electron heating scheme results in a relatively cold disk in this SANE simulation \citep[cf.][]{moscibrodzka2016,ressler2017}, causing the synchrotron emission to peak at higher latitudes where the temperatures are higher. 

\autoref{fig:photon_dist} shows the photon directions in the vicinity of the black hole as observed in the ZAMO frame. Here, the funnel region is defined as $\theta \geqslant 0.3\pi$, which is close to where $\sigma \geqslant 5$, and the disk region is taken to be within $0.1\pi$ of $\theta = \pi/2$. We report the angle between each photon's $k^{z}$ and $k^{R}$ components expressed in cylindrical coordinates in the ZAMO frame. Photons are emitted at small radius close to the equatorial plane ($\theta_{KS} \sim \pi/2$). In those locations, the photon distribution is nearly isotropic. Outside the emission region, photons preferentially travel outwards. Additionally, no photons are emitted from inside the funnel region where the MHD density and internal energy are unreliable. At small heights above the event horizon, photons are preferentially directed inwards towards the black hole as a result of light bending. This bending effect weakens with distance, resulting in an increasing fraction of outgoing photons with height until they are nearly all outgoing ($r \gtrsim 6 r_g$). SANE simulations seem less beamed in both the funnel and the equatorial plane in part because of the presence of two emission regions above and below the equatorial plane. \autoref{fig:photon_dist_vector} shows a schematic diagram illustrating the expected radiation field from compact, axisymmetric emission regions placed in the primary emission regions we have identified for the MAD W18 and SANE H10 simulations. The measured radial dependence of the radiation field anisotropy is consistent with the measured location of the emission region.

One promising site for gap formation is near the null surface around $r \simeq 2r_g$ \citep{chen2018,chen2020,kisaka2020}. At such small radii, the funnel radiation field is composed of nearly equal numbers of ingoing and outgoing photons. As a result, the net outward radiative flux is $F \ll 4 u_{\rm rad} / c$ and an analytic estimate of $u$ using the observed luminosity $L$ underestimates the photon number density and energy density by up to $1-2$ orders of magnitude.

\begin{figure*}
    \includegraphics[width=0.7\textwidth]{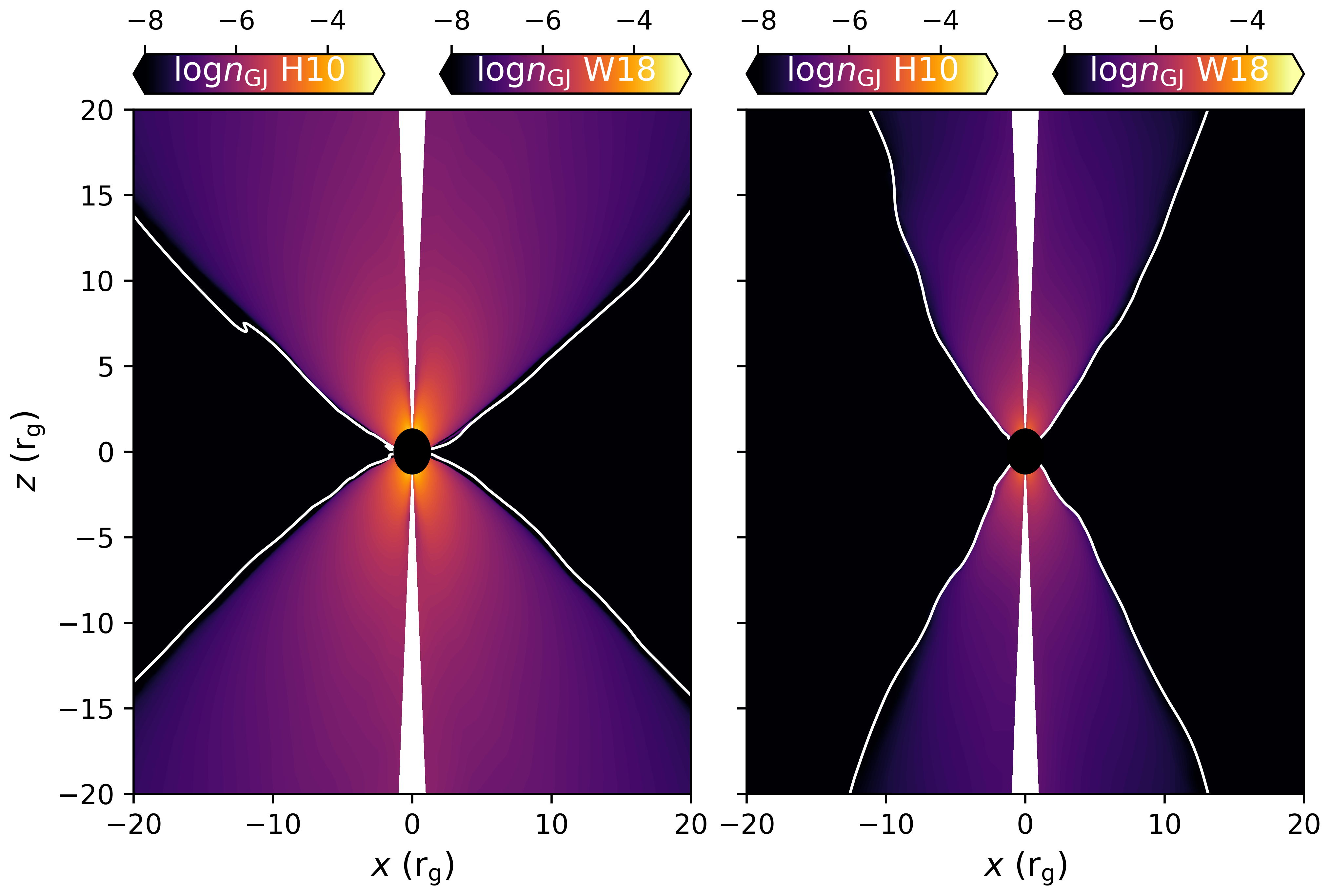}
    \caption{Log contour plots of funnel Goldreich--Julian density ($n_\mathrm{GJ}$) in units of $\rm cm^{-3}$ for MAD simulations (left) and SANE simulations (right) with different electron heating schemes. The white solid line marks the jet boundary defined as $\sigma \geqslant 5$.}
    \label{fig:nGJ}
\end{figure*}

\begin{figure*}
    \begin{tabular}{cc}
    \includegraphics[width=\columnwidth]{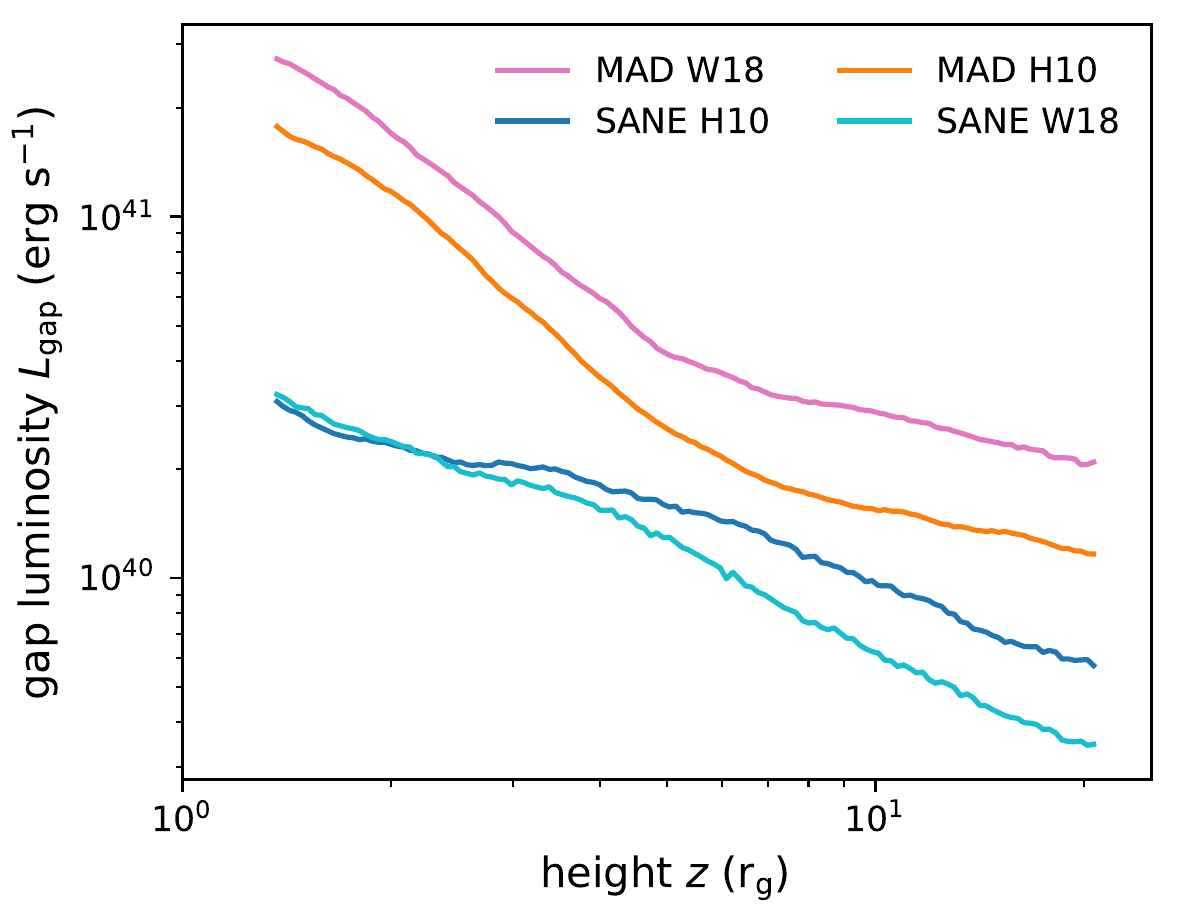} &
    \includegraphics[width=\columnwidth]{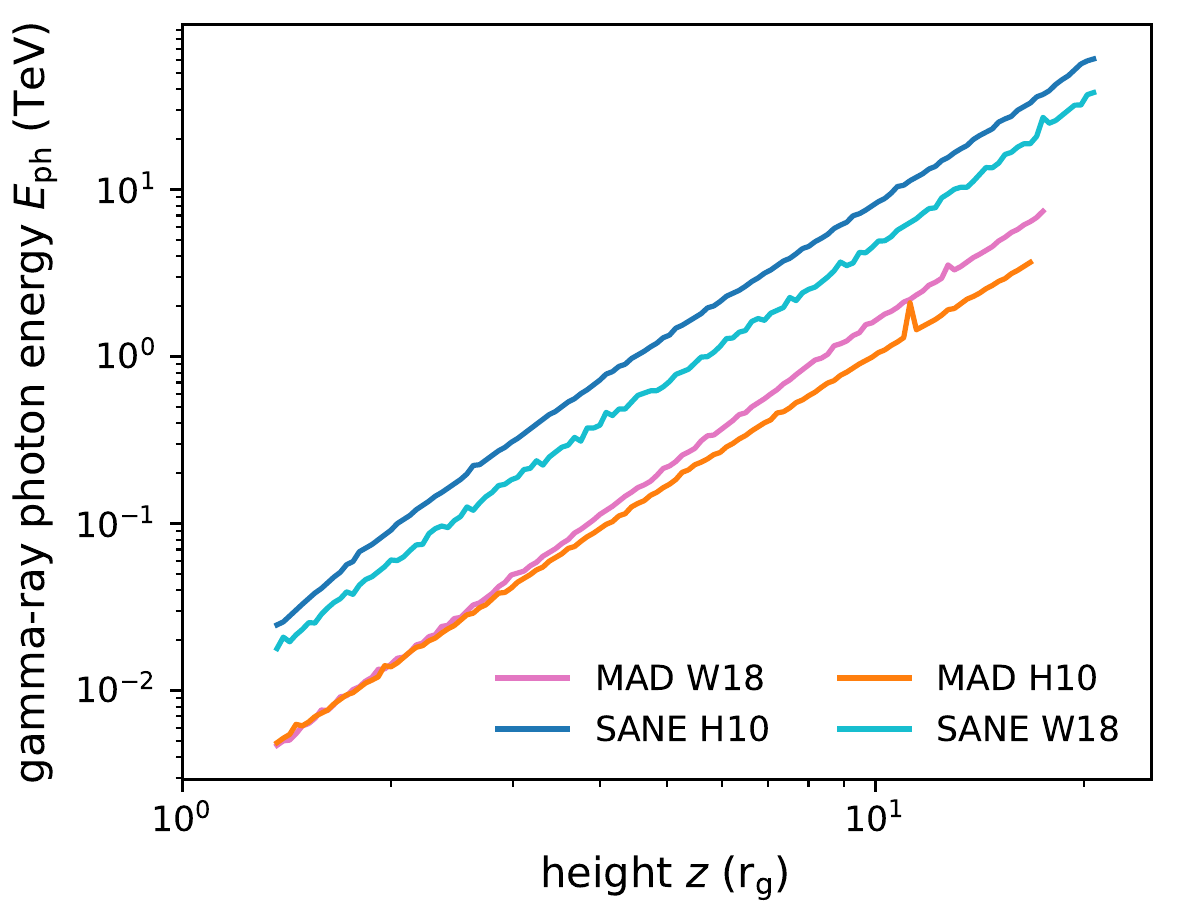}
    \end{tabular}
    \caption{Time-averaged gap luminosity (left) and maximum photon energy (right) as functions of radius for MAD simulation with W18 (pink) or H10 (orange) electron heating schemes and SANE simulation with W18 (cyan) or H10 (blue) schemes.}
    \label{fig:gap_luminosity_comp}
\end{figure*}

\begin{figure*}
    \begin{tabular}{cc}
    \includegraphics[width=\columnwidth]{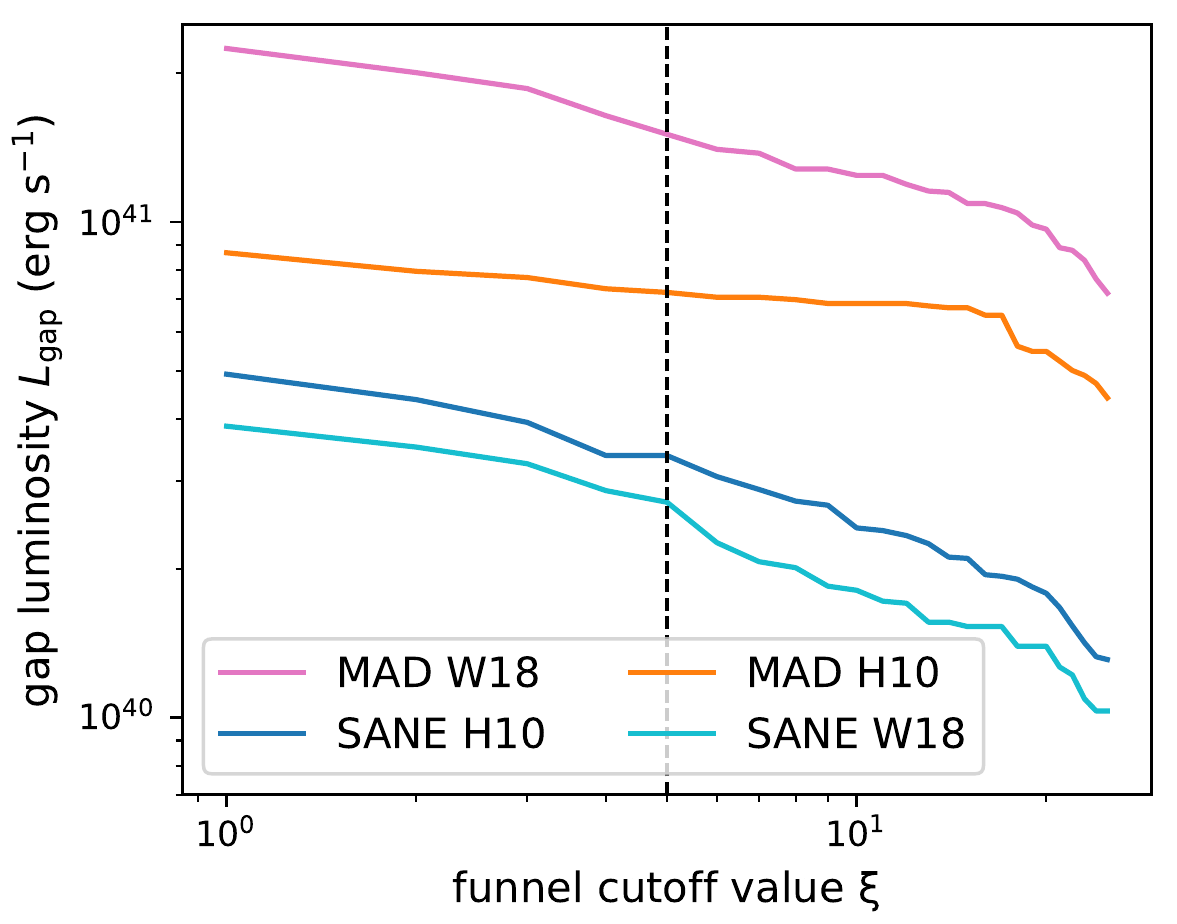} &
    \end{tabular}
    \caption{Time-averaged gap luminosity ($L_{\rm gap}$) near the null surface around $r \simeq 2r_g$ as a function of $\xi$, where $\sigma \geqslant \xi$ defines the funnel region. The dashed line marks our choice of $\xi = 5$ for defining the funnel boundary.}
    \label{fig:boundary}
\end{figure*}

\begin{figure*}
    \begin{tabular}{cc}
    \includegraphics[width=\columnwidth]{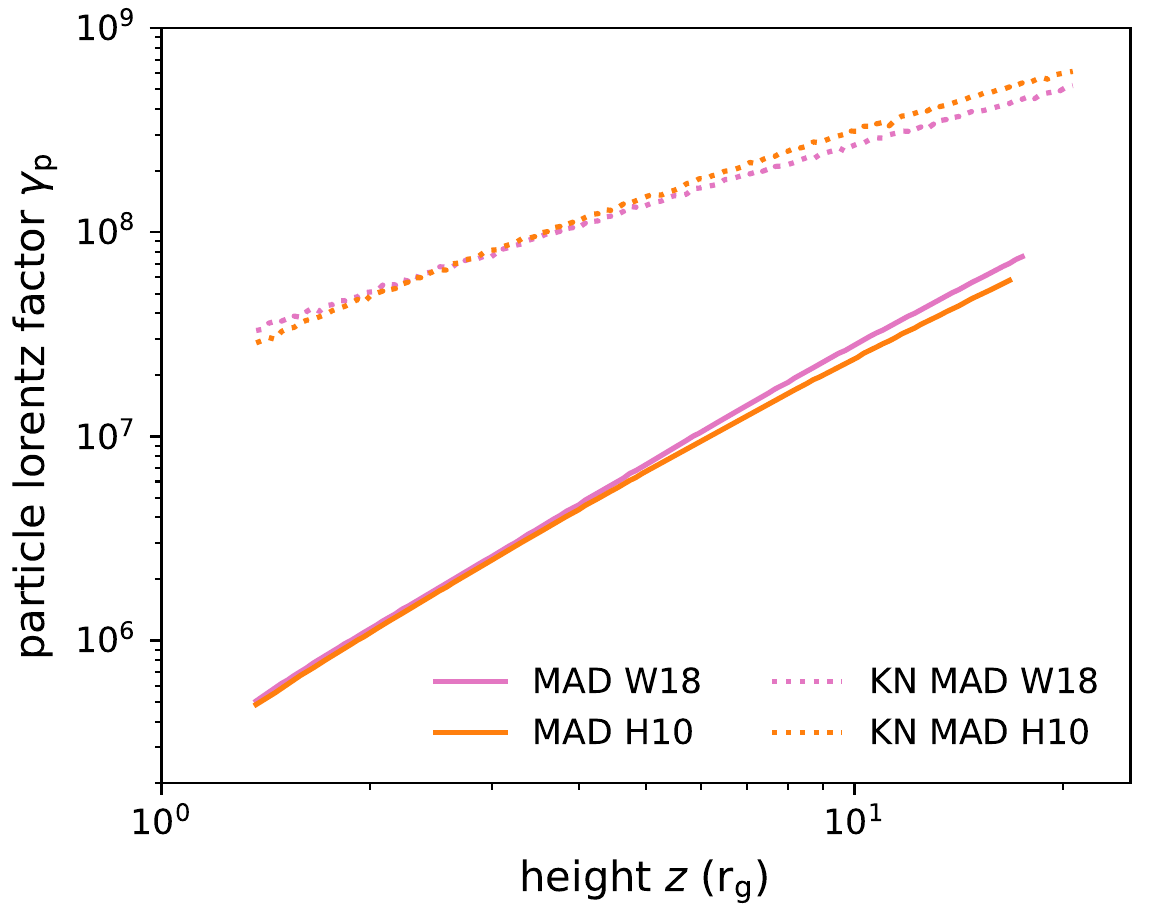}
    \includegraphics[width=\columnwidth]{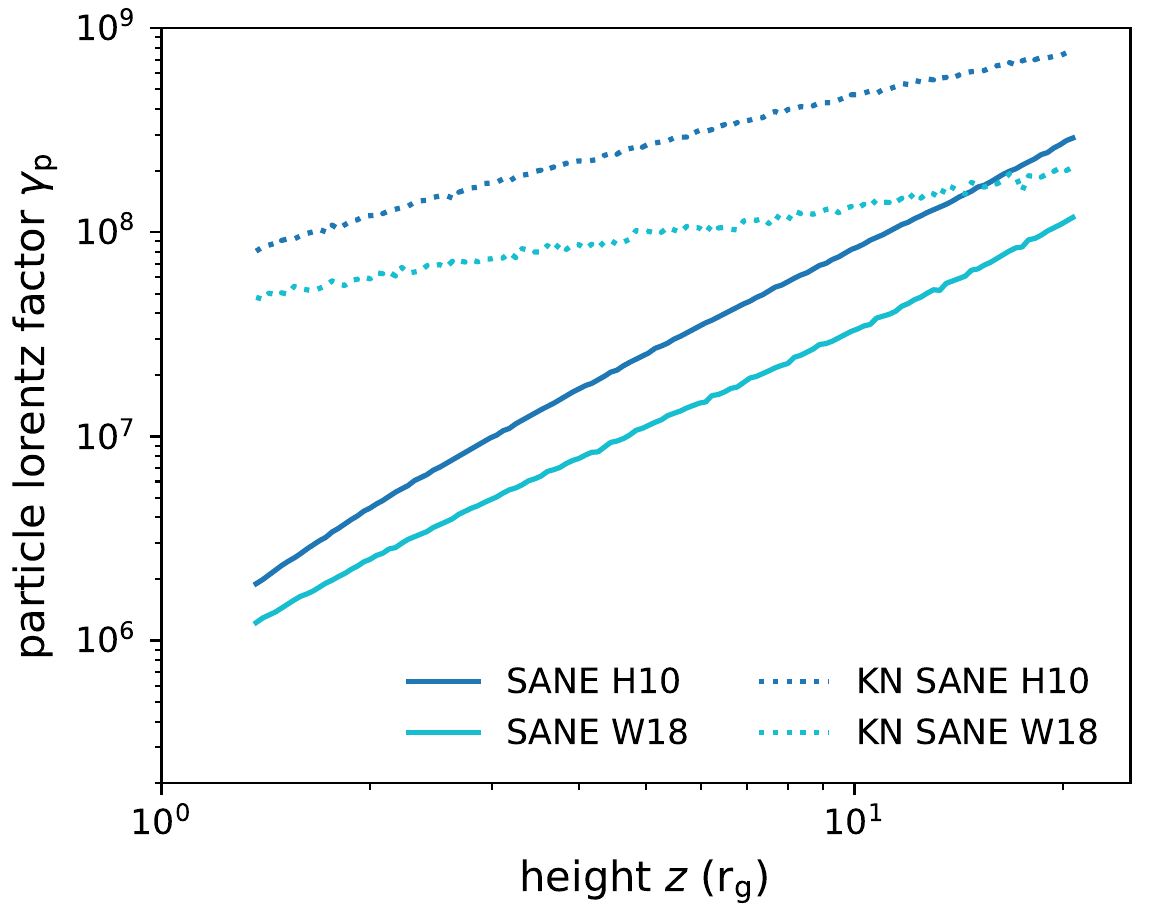}
    \end{tabular}
    \caption{Time-averaged Lorentz factor (solid line) and associated Klein--Nishina limit $\gamma_p \sim 0.1/\epsilon_\mathrm{min}$ (dotted line) are plotted against radius. MAD simulations with W18 (pink) and H10 (orange) schemes are on the left panel; SANE simulations with W18 (cyan) and H10 (blue) schemes are on the right panel}
    \label{fig:particle_lorentz_factor}
\end{figure*}

\begin{table*}
    \centering
    \caption{Characteristic parameters of four sets of simulation results averaged between 2 and 3  $r_g$.}
    \label{tab:results_comp}
    \begin{tabular}{lccccc}
    \hline
Parameters & MAD W18 & MAD H10 & SANE W18 & SANE H10\\
\hline
$n_s \, (\rm cm^{-3})$ & $3.7\times 10^{14}$ &     $3.6\times10^{14}$ & $1.2\times10^{14}$ & $8.6\times10^{13}$\\
$l_{\rm IC}$ & $4.6\times10^{9}$ &  $5.0\times10^{9}$ & $1.5\times10^{10}$ & $2.0\times10^{10}$\\
$\tau_0$ & $2.3\times10^{5}$ & $2.2\times10^{5}$ & $7.0\times10^{4}$ & $5.3\times10^{4}$\\
$u_{\rm rad} \, (\rm erg \hspace{0.1cm}  cm^{-3})$ &   2.9 &   3.4 & 0.78 & 0.30\\
$\epsilon_{\rm min}/ {m_e}c^2$ &    $1.8\times10^{-9}$ &    $1.9\times10^{-9}$ & $1.5\times10^{-9}$ & $7.2\times10^{-10}$\\
$\epsilon_{\rm min} \, (\rm erg)$ &    $1.4\times10^{-15}$ &   $1.5\times10^{-15}$ & $1.2\times10^{-15}$ &$5.9\times10^{-16}$\\
$j_B$ (cgs) &   $3.3\times10^{-4}$ &    $2.6\times10^{-4}$ & $1.2\times10^{-4}$ &$1.1\times10^{-4}$\\
B (G) &  210 & 150 & 52 & 44\\
$E_0 \, (\rm statV \hspace{0.1cm} cm^{-1})$ &  $1.5\times10^{-5}$ &    $1.3\times10^{-5}$ & $9.0\times10^{-6}$ & $8.7\times10^{-6}$\\
$\lambda_p/\rm r_g$ & $ 1.3\times 10^{-7}$ & $1.5\times 10^{-7}$ & $2.2\times 10^{-7}$ & $2.3\times 10^{-7}$\\
$\gamma_p$ & $1.5\times10^6$ & $1.5\times10^6$ & $2.9\times10^6$ & $6.4\times10^6$\\
$E_{\rm ||}/E_0$ &  750 &   780 & 1300 & 2200\\
$E_{\rm ||} \, (\rm statV \hspace{0.1cm} cm^{-1})$ &   0.011 &     0.010 & 0.011 & 0.019\\
$L_{\rm gap} \, (\rm erg \hspace{0.1cm} s^{-1})$ (1$r_g$ gap size) &   $1.3\times10^{41}$ &    $8.6\times10^{40}$ & $2.1\times10^{40}$ & $2.2\times10^{40}$\\
\hline
    \end{tabular}
    
\end{table*}

\section{Funnel particle density and prospects for gap formation}
\label{sec:funnel_particles}

In the previous section we studied the radiation and magnetic field properties of the funnel region, which can be reliably measured using radiation GRMHD simulations. The matter content inside the funnel, however, is dominated by artificially injected mass and energy (``floors") needed for a stable evolution of the code. In reality, a physical mechanism is required to supply the plasma needed in the jet funnel. Here we discuss two physical scenarios for particle loading in the funnel region and their implications for particle acceleration.

\begin{figure}
    \includegraphics[width=0.85\columnwidth]{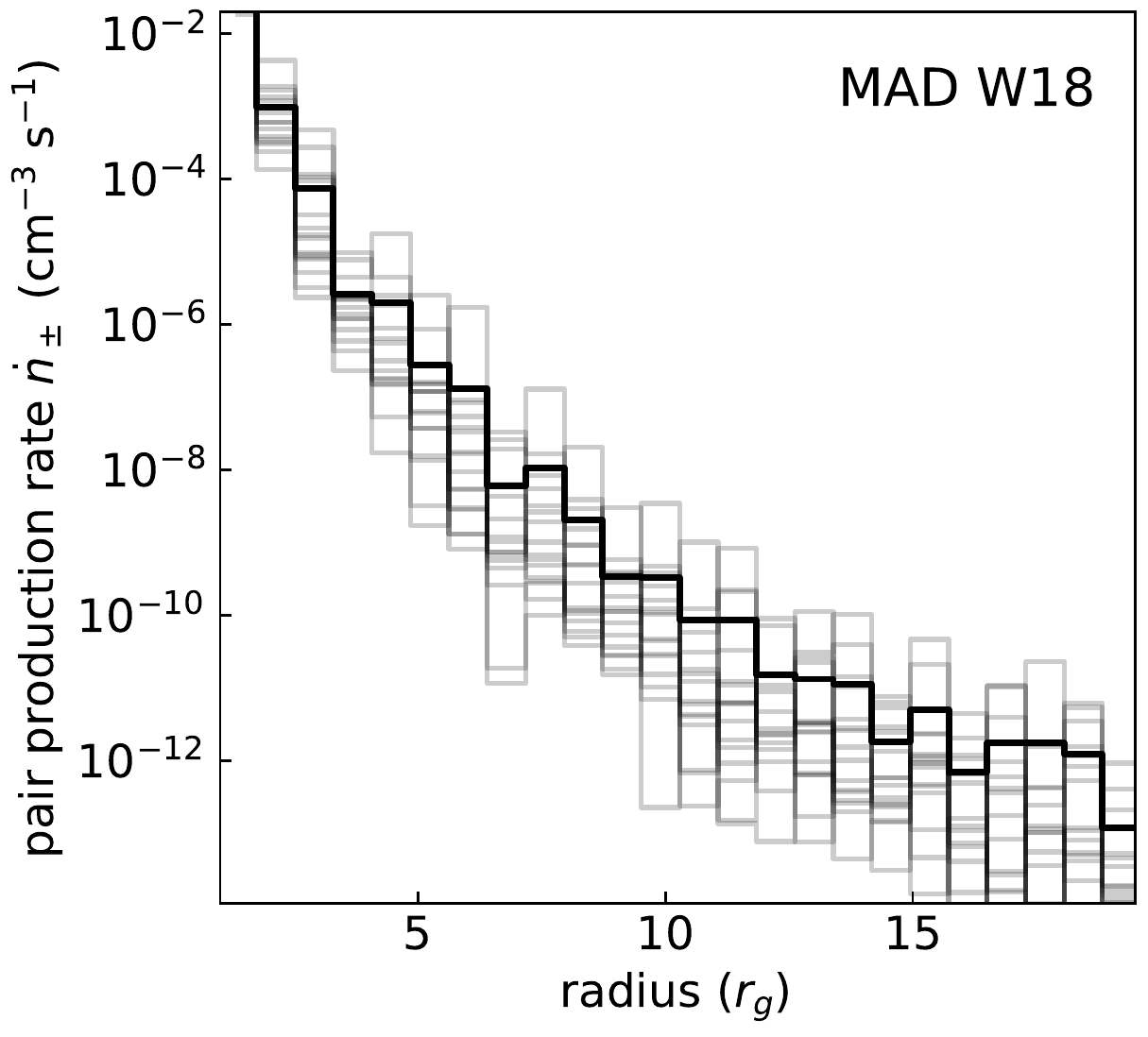}
    \caption{Photon--photon pair-production rate as a function of height at one pole ($\theta \lesssim 5^\circ$) for the MAD W18 simulation. The thin gray lines show the result from the photon distribution in azimuthal wedges of $\Delta \phi = 22.5^\circ$ at a snapshot in time, and the thick black line is their average.}
    \label{fig:drizzlerate}
\end{figure}

\subsection{Pair production from annihilation of photons produced in the accretion flow}
\label{sec:drizzle}

Photons produced by the background radiation field of the accretion flow may interact with one another and convert
into electron--positron pairs. This ``drizzle'' pair creation process provides a mechanism to mass-load the otherwise evacuated funnel. Photons with energies $h\nu > m_e c^2$ ($\gtrsim 1$ MeV) are produced in radiation GRMHD simulations by inverse Compton scattering. Post-processing calculations of the pair production process can provide quantitative estimates of the funnel particle density \citep[e.g.,][]{moscibrodzka2011,wong2020}. 

The required minimum charge density to screen the electric is the Goldreich-Julian density \citep[e.g.,][]{wong2020}:
\begin{equation}
n_\mathrm{GJ} = {-u^{\mu}j_{\mu} \over e c^2},
\end{equation}
\noindent where $u^{\mu}$ is the four-velocity of the fluid and $j_{\mu}$ is the four-current.
\autoref{fig:nGJ} shows azimtuhally averaged maps of $n_{\rm GJ}$ inside the funnel. The maximum values of $\simeq 10^{-4} \, \rm cm^{-3}$ are $\simeq 8$ orders of magnitude lower than particle densities in the accretion flow. This corresponds to an electron magnetization parameter $\sigma_\mathrm{GJ}\sim 10^{13}$.

Following \citet{moscibrodzka2011} and \citet{wong2020}, we estimated the pair-production rate from drizzle. From \textsc{ebhlight} superphoton four-momenta $k^\mu$ recorded at a single snapshot in time, the pair production rate is estimated as,

\begin{equation}
    \dot{n}_\pm = \frac{1}{2} \sum_{ij} \frac{w_i w_j}{(\sqrt{-g} \hspace{4pt} \Delta^3 x)^2} \frac{k_i^\mu k_{j, \mu}}{k_{0,j} \, k_{0,i}} \sigma_{\gamma \gamma} \, c,
\end{equation}

\noindent where $\sigma_{\gamma \gamma}$ is the pair-production cross section \citep[e.g.,][]{breit1934}, $w_i$ is the number of photons within each Monte Carlo superphoton sample, and the sum is performed over all pairs of superphotons. For computational efficiency, we take the sum $i$ to be only over photons exceeding the pair threshold, with energies $> 0.5 m_e c^2$, while the sum $j$ is taken over all photons. The pair drizzle calculation is performed using a short, dedicated \texttt{ebhlight} calculation with $\sim 10^{10}$ superphoton packets in order to provide a resolved sample of the high energy radiation field. We carry out the calculations at one snapshot in time using a grid distinct from that used for the fluid, consisting of $24 \times 32 \times 16$ zones evenly spaced in $r$, $\theta$, and $\phi$ in Kerr-Schild coordinates. The values of $\sqrt{-g}$ and $\Delta x$ above refer to those calculated at the centers of these new zones and to their total sizes.

\autoref{fig:drizzlerate} shows the result near the pole for the MAD W18 model, where $\dot{n} \, r_g / c \sim 0.1$ cm$^{-3}$ at $r \simeq 3 \, r_g$. For $n_\pm \sim \dot{n} \,  r_g/c$, these results imply $n_\pm / n_{\rm GJ} \gtrsim 10^{3-4}$, which is sufficient to screen electric fields and prevent gap formation. The steep decline in pair-production rate with radius also matches the results of previous work \citep{moscibrodzka2011}. \citet{wong2020} further derived a fitting formula where $\dot{n}_\pm \sim \dot{m}^5$ for our physical regime of interest. This suggests that a change in mass accretion rate by factors of $\gtrsim 5$ could be sufficient to trigger gap formation in the MAD W18 model. Typically, $\dot{M}$ varies by factors of $\simeq 2-3$ in GRMHD simulations.

\subsection{Luminosity estimate in case of gap formation}
\label{sec:lgap}

Based on the estimates above, sufficient charge carriers are likely available to screen electric fields in the funnel region. Since a moderate change to the accretion flow or jet could plausibly result in regions with  $n_\pm / n_{\rm GJ} < 1$, we estimate the $\gamma$-ray luminosity from a gap that forms under such conditions.  \citet{chen2018} used time-dependent 1D PIC simulations to study particle acceleration due to gap formation in a black hole magnetosphere. They found that quasi-periodic operation of a gap in M87's near-horizon funnel region can produce TeV flares with intermittency similar to the observed variability timescale. 
Drawing empirically from their simulation results, they developed an analytic model to estimate the parallel electric field strength and in turn the gap luminosity. Here, we will first recapitulate the simple analytic model, then use the physical quantities computed from our radiation GRMHD simulations as inputs to the model, considering the SANE and MAD accretion flows along with the H10 and W18 electron heating models. 

Assuming charge carriers with number density $n$ move at close to the speed of light, we can approximate the background current density $j_B = enc$, which aids the estimation of the plasma frequency $\omega_p$ and the plasma skin depth $\lambda_p$ as follows:
\begin{equation}
\omega_p = \sqrt{\frac{4\pi n e^2}{m_e}} = \sqrt{\frac{4\pi e j_B}{m_ec}}, \quad \lambda_p = \frac{c}{\omega_p}.
\end{equation}

Requiring each electron in $E_0$ to gain energy equal to $m_ec^2$ after accelerating over a distance of $\lambda_p$, this electric field unit is defined as:
\begin{equation}
E_0 = \frac{m_ec^2}{e\lambda_p} = \sqrt {4\pi j_B m_e c \over e}.\\
\end{equation}

Scattering with background soft photons generates $\gamma$-rays with $\kappa$ of them converted to pairs within time t:
\begin{equation}
    \kappa = {c^2 t^2 \over 5\ell_{\rm IC}^2} (\gamma_p^2\epsilon_{\rm min}^2)^\alpha \sim 10,
\end{equation}
where $\alpha \sim 1.2$ is the spectral index, $\gamma_p$ is the primary particle Lorentz factor, and $\epsilon_{\rm min}$ is the minimum photon energy in units of $m_ec^2$. For the calculation of $L_{\rm gap}$, we use an empirical power-law seed photon spectrum rather than the spectral shape found from our radiation GRMHD simulations. This choice is made both for consistency with the work of \citet{chen2018}, whose 1D GR PIC simulations assume a power law distribution of seed photons, and because our current models do not accurately reproduce the infrared to X-ray SED shape observed from M87. New GR PIC calculations based on photon distributions from radiation GRMHD may find different pair cascade properties and gap luminosities.

For a gap height of $\sim 1 \, r_g$,the gap electric field $E_{\rm ||}$ in units of $E_0$ can then be calculated as:
\begin{equation}
{E_{||} \over E_0} = \left({5 \kappa \ell_{\rm IC}^2 \over 8 \pi \lambda_p r}\left({{3 (\alpha - 1)} \over 4 \alpha} {\ell_{\rm IC} \epsilon_{\rm min} \over \lambda_p}\right)^{-\alpha}\right)^{1 \over {\alpha+1}},
\end{equation}
where $\ell_{\rm IC} = 1/n_s\sigma_T$ is the mean free path for an electron to undergo inverse Compton scattering in the Thomson limit.

We assume a constant gap size $h$, and allow the gap to appear anywhere in the funnel between the horizon radius of 1.37 $r_g$ and $\sim 40r_g$, where the photons are collected. We then calculate the gap luminosity as a function of its height above the black hole $z$ as:
\begin{equation}
L_{\rm gap}(z) \simeq {\pi \Delta x_2 \, r_g^3 \, {h \over r} \sum_k {\sqrt{-g}E_{||,k}j_{B,k}}}.
\end{equation}
Here, all variables are $\phi$-averaged, and $\Delta x_2$ is related to the polar angle spacing in our simulation grid. In addition, the summation is done across the angular region at a given radius inside which $\sigma \geqslant 5$. 

The Lorentz factors of accelerated particles are limited by inverse Compton cooling, therefore they are determined by the acceleration $E_\parallel$ and the scattering free path:
\begin{equation}
\gamma_p^2 = {{{3eE_{||} (\alpha - 1) \ell_{\rm IC}}} \over {4 \alpha \epsilon_{\rm min}m_e c^2}}.
\end{equation}
The model described above works well when inverse Compton scattering occurs mostly in the Thomson regime, when $\gamma_p \lesssim 0.1/\epsilon_\mathrm{min}$\citep{chen2018}. In this regime, the scattered photon energy can be computed simply as:
\begin{equation}
E_{\rm ph} = 2\gamma_p^2\epsilon.
\end{equation}

We use the photon number density, radiation energy density, and current density from the simulation as inputs into the analytic model. All data are averaged over the final $\sim 2000 \, r_g/c$ of each simulation. \autoref{fig:gap_luminosity_comp} shows the gap luminosity produced by the parallel electric field for a gap of size $h = 1 r_g$ as a function of the height $z$ where it appears. MADs have a gap luminosity around $\rm 10^{41}\, erg\, s^{-1}$ in contrast to the value around $\rm 2 \times 10^{40}\, erg\, s^{-1}$ for SANEs. The difference can be partially attributed to the higher product of $E_{||}$ and $j_B$ and partially to the larger funnel size seen in MAD simulations. Varying the electron heating scheme results in much smaller differences. All four simulations show a decreasing trend of gap luminosity with distance from the horizon. \autoref{tab:results_comp} gives typical values of each simulation variable averaged over azimuthal and polar angle, between 2 to 3 $r_g$, and inside the defined funnel region. The choice of the critical value of $\sigma$ to define the funnel boundary ($\xi$) leads to modest changes in the gap luminosity (\autoref{fig:boundary}).

All simulations are capable of producing TeV $\gamma$-ray photons depending on the radius at which the gap electric field emerges. As presented in \autoref{fig:gap_luminosity_comp}, all trends suggest that for a gap appearing further away from the horizon, the produced gap luminosity decays while $\gamma$-ray photon energy increases, reaching TeV range ($>0.3\,\mathrm{TeV}$) at approximately 3 $r_g$ for SANE simulations and 6 $r_g$ for MAD simulations. The Lorentz factors to boost to high-energy shown in \autoref{fig:particle_lorentz_factor} are all below the corresponding Klein-Nishina limit where \citet{chen2018}'s analytic model applies. However, we do not see significant flares in the time variability with only a dozen samplings. The simulation duration is limited by the computational expense of 3D radiation GRMHD.

\section{Discussion}
\label{sec:discussion}

We have presented 3D radiation GRMHD simulations of accretion onto the supermassive black hole in M87. The simulations were performed with the public code \textsc{ebhlight}, which evolves the frequency-dependent radiation field using a Monte Carlo method. We have explored regimes of weak and dynamically important accumulated magnetic flux (SANE and MAD), as well as different sub-grid prescriptions for electron heating from either turbulence (as realized in gyrokinetics calculations) or reconnection (as realized in PIC calculations). Radiative cooling lowers the average electron temperature by factors of $\simeq 2-5$. For SANE models, our $T_e$ maps are consistent with those from the  axisymmetric \texttt{ebhlight} calculations of  \citet{ryan2018}. MAD models using the H10 prescription show higher $T_e$ and much stronger radiation pressure in the funnel than when using the W18 or R17 prescriptions. This finding agrees with that of \citet{chael2019}, and is even more pronounced in a short MAD H10 run where we included radiation for $\sigma < 25$. However, those changes do not seem to play a major role in influencing the accretion flow dynamics \citep[but see][]{chael2019}. Spectra from synchrotron radiation and inverse Compton scattering from a purely thermal electron distribution can reproduce the radio/submillimeter and in some cases X-ray luminosity of M87. 

The choice of electron heating prescription in combination with the SANE or MAD state leads to different electron temperature distributions, and in turn the relative importance of inverse Compton scattering (e.g., the $y$-parameter). The gyrokinetic turbulence prescriptions (H10 and K19) produce colder electrons in the accretion flow body, resulting in lower scattered luminosities especially in the SANE case. The high density and disk electron temperature in the parameter combinations of SANE with reconnection heating (W18 and R17) result in the models systematically overproducing the observed X-ray luminosity. The MAD models are already radiatively efficient at low luminosity ($L_{\rm bol}/\dot{M}c^2 \approx 10\%$), while the efficiencies of the SANE models are roughly an order of magnitude smaller. All radiative efficiencies are about an order of magnitude larger than in similar, non-radiative simulations of Sgr A* \citep{dexter2020harmpi}.

We use the simulations to study the radiation and magnetic field properties in the evacuated, jet funnel region. The funnel is systematically wider in the MAD case, consistent with past results. The radiation field is generally anisotropic, and the relative preferred photon direction varies systematically with height above the black hole from ingoing near the horizon to outgoing for $r \gtrsim 6 \, r_g$. This result can be explained in terms of the compact, near equatorial emission regions in our simulations. 

By comparing a rough estimate of the funnel pair-production rate with the Goldreich--Julian density, it appears that the charge density in the funnel may be high enough to screen the electric fields. However, the strong dependence of the pair-production rate with mass accretion rate suggests that a moderate change in accretion flow properties could in principle trigger gap formation and particle acceleration in the funnel. We use  radiation and magnetic field quantities from our simulations (\autoref{tab:results_comp}) to estimate $L_{\rm gap}$ using \citet{chen2018}'s models, assuming a gap is able to form:
\begin{equation}
L_{\rm gap} \sim E_{||} j_{\rm B} r_g^3 \sim {E_{||} \over B}L_{\rm jet},
\end{equation}
which provides a consistent estimate of $L_{\rm gap}$ for a jet power of $\sim 10^{43}\,\mathrm{erg/s}$. The resulting gap luminosity varies between $10^{40}\,\mathrm{erg/s}$ to a few times $10^{41}\,\mathrm{erg/s}$ depending on the electron heating scheme, with SANE simulations providing much closer values to \citet{chen2018}'s estimates of $\sim 2\times10^{40}\,\mathrm{erg/s}$. This fits most of the predictions from various simulation processes and assumptions \citep{rieger2020,chen2020}, as well as reported observational evidence of M87's VHE flares with  $L_{\rm VHE} \simeq (0.8-2.4) \times 10^{42}\,\mathrm{erg/s}$ \citep{abramowski2012}. Compared with the earlier work by \citet{2015ApJ...809...97B}, the PIC models that we use generally predict lower total gap luminosity, especially at high optical depth $\tau_0$. This is because pair production through photon collision becomes very efficient and the gap is quickly screened. \citet{2015ApJ...809...97B} found that the gamma-ray luminosity from the gap can be very high ($\sim 10^{43}\,\mathrm{erg/s}$), but most of it is beamed away from us. Our model does not require this assumption, but a future more careful examination of radiation beaming in the presence of the magnetic field structure from GRMHD simulations may be useful to further constrain the observed gamma-ray luminosity.

Recently \citet{2020ApJ...902...80K} also performed 1D GR PIC simulations of gaps. They focused on the regime where the fiducial optical depth $\tau_{0}$ is between 10 to 300 and concluded that the luminosity of curvature radiation from the accelerated particles can dominate the high energy emission in some models. In this work, we have found that the optical depth in the vicinity of M87 can be much higher, $\tau_{0} \sim 10^{5}$. In this regime, inverse Compton cooling is much more efficient, limiting the particles to lower Lorentz factors. Since the luminosity of curvature radiation scales as $\gamma^{4}$, it becomes subdominant in producing high energy gamma-rays.

We have assumed a uniform gap size of $h = 1 r_{g}$ in our estimates. The gap size likely has a weak dependency on the optical depth $\tau_{0}$ and the spectrum of the soft photons. A recent work by \citet{crinquand2020} showed that in an axisymmetric configuration, under a mono-energetic soft photon background, the gap can be very narrow, $h \sim 0.05r_{g}$. A more systematic study to understand the scaling of the gap size with geometry, optical depth, and the background photon distribution will help to improve upon our estimates for the gap luminosity.

We only see a small variability of the gap luminosity from our simulations by factors of $\lesssim 2$, which is insufficient to explain the large amplitude of VHE flaring observed from M87. Variability properties would be interesting to study in longer duration radiation GRMHD simulations with a higher dump cadence. Our preliminary results suggest that flares might instead require a large-scale change to the funnel structure or jet power, or alternatively could be triggered by small-scale kinetic processes within the funnel region itself.


\section*{Acknowledgements}
We thank M. C. Begelman, C. F. Gammie, A. Philippov, A. Tchekhovskoy, and D. Uzdensky for helpful discussions, and the referee, A. Chael, for constructive comments which improved the paper. This work was supported in part by NASA Astrophysics Theory Program grant 80NSSC20K0527, an Alfred P. Sloan Research Fellowship, a Sofja Kovalevskaja award from the Alexander von Humboldt foundation, a Donald
C. and F. Shirley Jones Fellowship, and the Undergraduate Research Opportunities Program (UROP) at the University of Colorado Boulder. The calculations presented here were carried out on the MPG supercomputer Cobra hosted at MPCDF and used resources supported by the NASA High-End Computing (HEC) Program through the NASA Advanced Supercomputing (NAS) Division at Ames Research Center. This work has been assigned a document release number LA-UR-21-21285.

\section*{Data Availability}
The simulation data analyzed in this article will be shared on reasonable request to the corresponding author.



\bibliographystyle{mnras}
\bibliography{ebhlight} 




\appendix

%

\section{Convergence tests}
\label{app:nph_converge}

\begin{figure*}
\begin{tabular}{cc}
    \includegraphics[width=0.88\columnwidth]{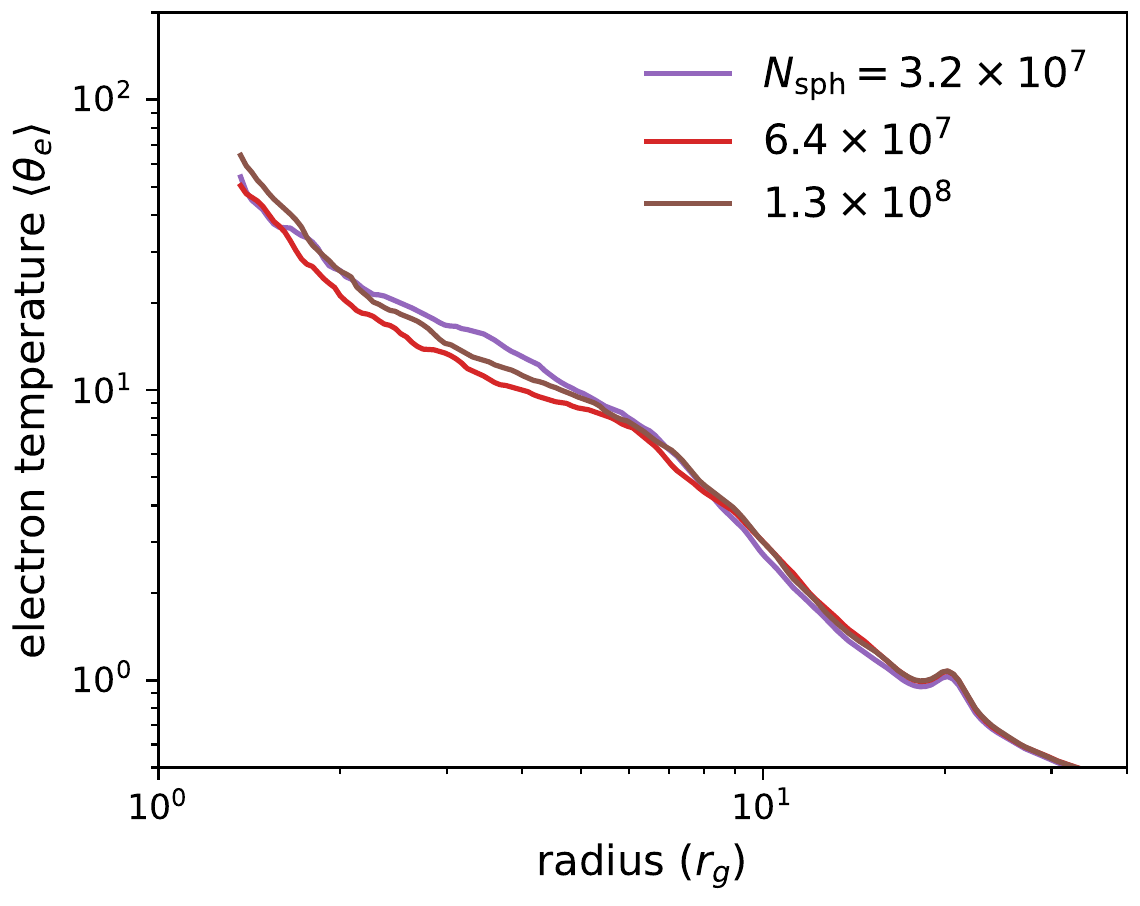} &
        \includegraphics[width=0.9\columnwidth]{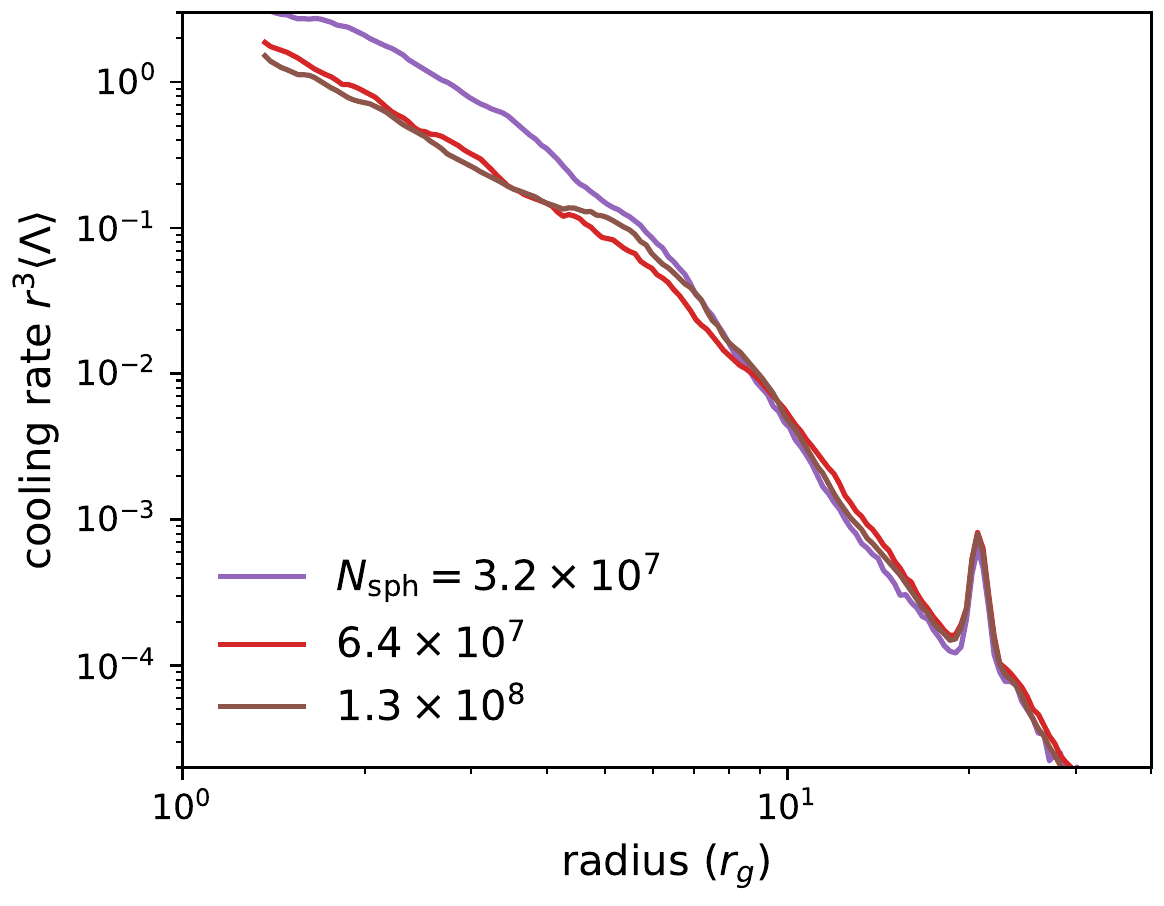}\\
    \includegraphics[width=0.88\columnwidth]{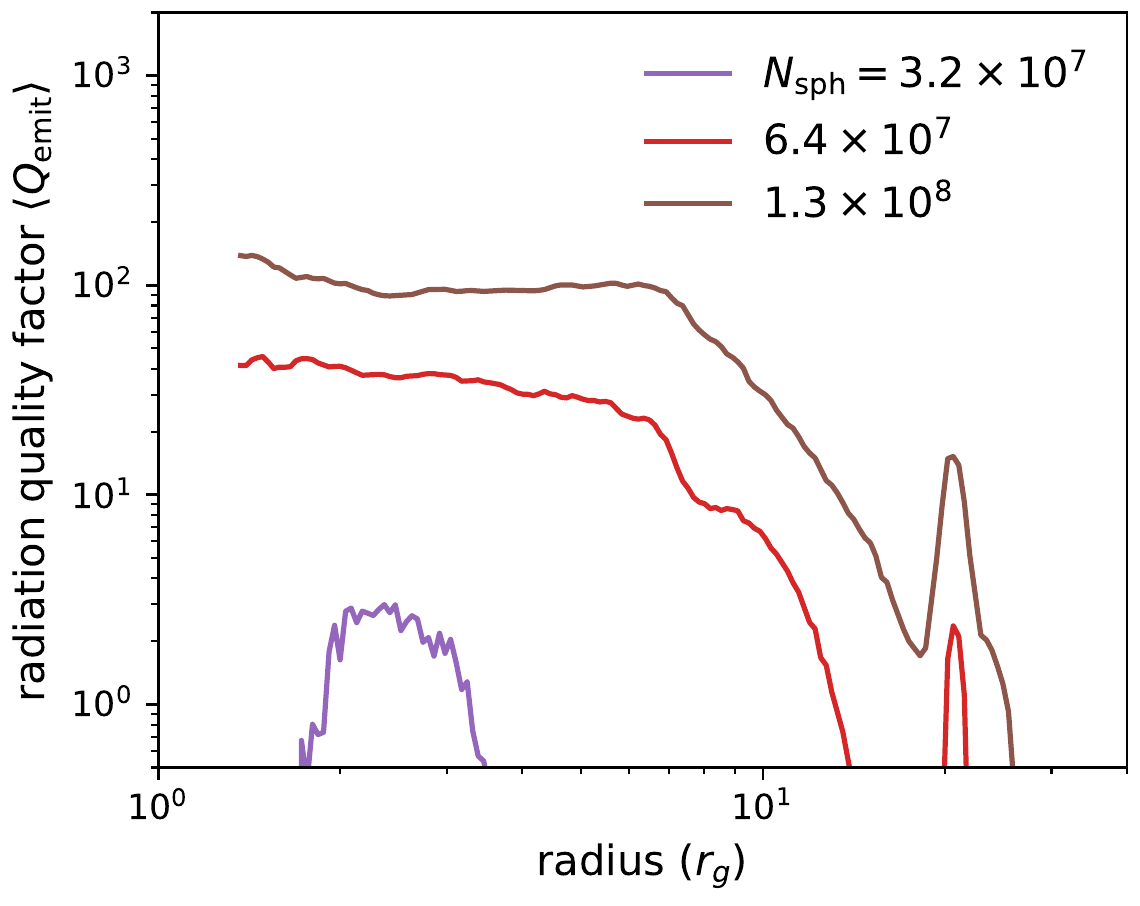} &
    \includegraphics[width=\columnwidth]{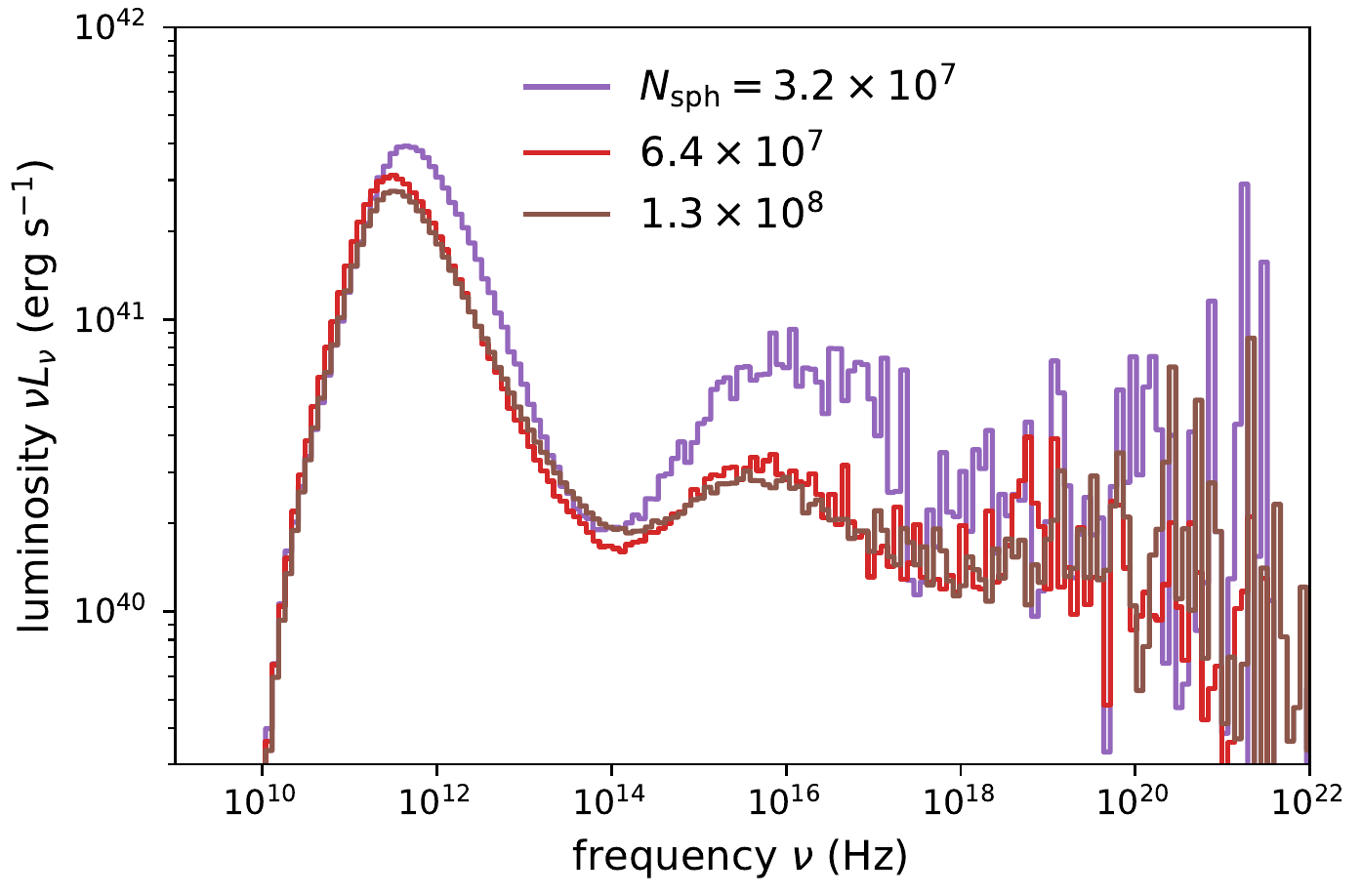}
    \end{tabular}
    \caption{Electron temperature (top left), integrated cooling rate (top right), emission quality factor (bottom left), and time-averaged SEDs (right) as a function of the number of superphotons used for the MAD H10 model. Our fiducial resolution of $6.4 \times 10^7$ superphotons ($\simeq 5-10 \times$ the number of fluid grid zones) gives similar time-averaged results as using twice that number. The radiation quantities begin to diverge at half the fiducial number. The quality of the radiation solution seems well captured in this case by the quality factor $\langle Q_{\rm emit} \rangle$.}
    \label{fig:restest_rad}
\end{figure*}

\begin{figure*}
\begin{tabular}{cc}
    \includegraphics[width=0.88\columnwidth]{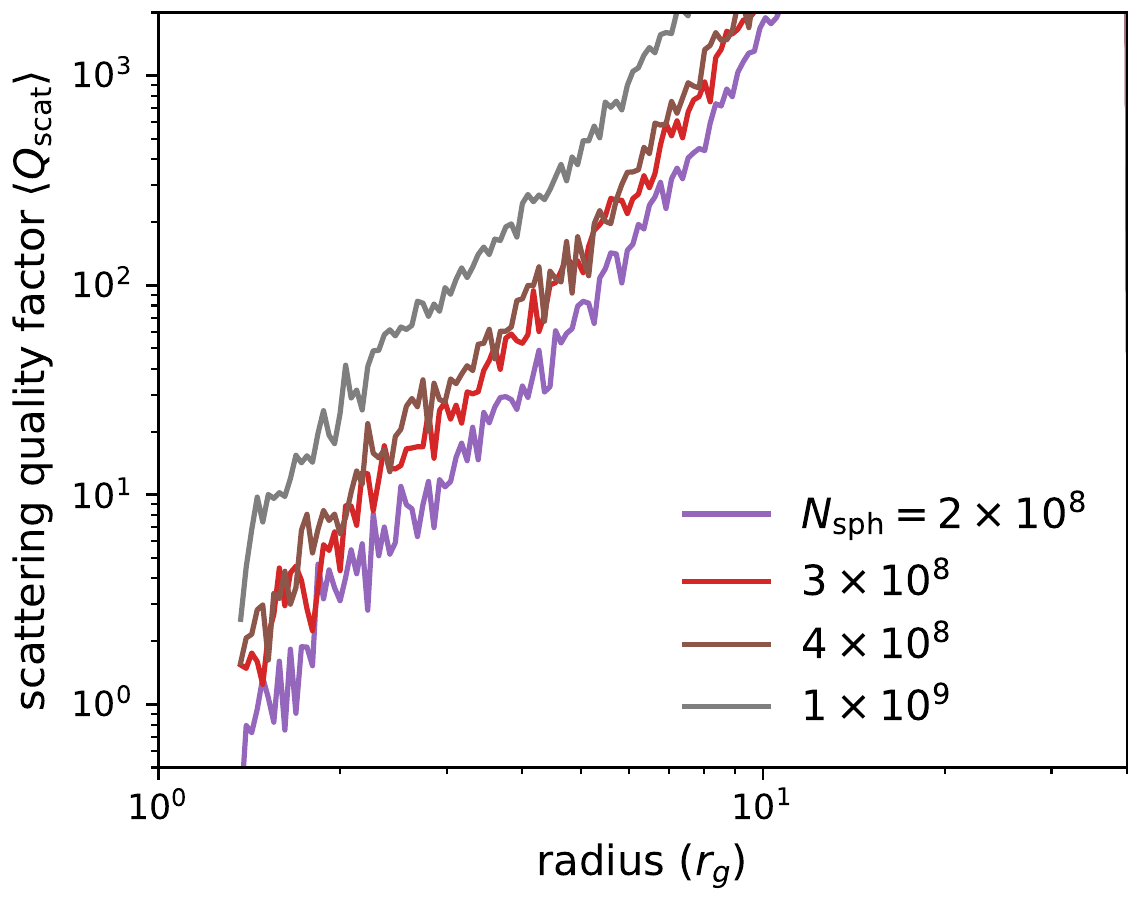} &
     \includegraphics[width=0.88\columnwidth]{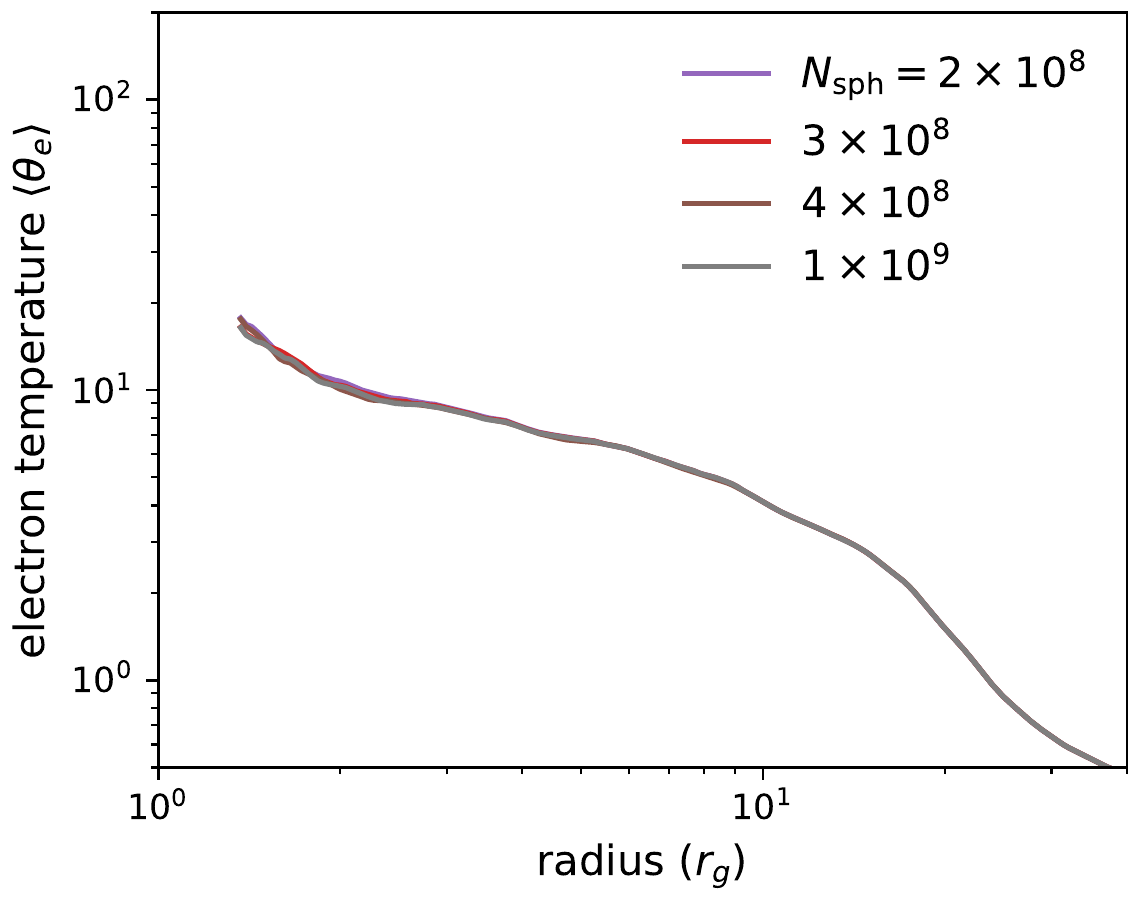}\\
\end{tabular}
    \caption{Scattering quality factor (left) and electron temperature (right) as a function of the number of superphotons used for the MAD W18 simulation. The numbers of superphotons used for this test are $\simeq 3-15\times$ larger than in our fiducial simulations. The scattering quality factor increases with superphoton resolution, and the electron temperature (and other radiation field dependent variables), appear well converged. This suggests that our low values of $\langle Q_{\rm scat} \rangle$ may not severely compromise the accuracy of the radiation field solution very close to the horizon ($r \lesssim 3 \, r_g$.)}
    \label{fig:restest_rad_W18}
\end{figure*}

To test for convergence of the radiation field, we compare shell-averaged radial profiles of various radiation quantities using different average numbers of superphotons in our MAD H10 simulation (\autoref{fig:restest_rad}). We expect a converged solution when a sufficient number of superphotons are used in the calculation. The solutions are nearly identical between our fiducial resolution and twice the number of superphotons. Using half the number results in larger  radiation pressures and integrated cooling rates. The \textsc{ebhlight} radiation quality factor $\langle Q_{\rm emit} \rangle$ is $\gtrsim 10$ for the fiducial case and $\gtrsim 30$ when using twice the number of superphotons. It falls below $1$ in the lower resolution case. Based on our results, it seems that values  $\langle Q_{\rm emit} \rangle \lesssim 10$ may indicate insufficient superphoton resolution.

For all simulations presented here, the \emph{scattering} quality factor $\langle Q_{\rm scat} \rangle$ is $\ll 10$ for small radii $r \lesssim 3 \, r_g$. This indicates that the inverse Compton cooling may be inaccurate near the event horizon. Since $L_{\rm scat} \simeq L_{\rm emit}$ in many of our simulations (\autoref{tab:rad_table}), a large value of $\langle Q_{\rm emit} \rangle$ may be insufficient to assess convergence. \autoref{fig:restest_rad_W18} shows the results of running the MAD W18 simulation (where $L_{\rm scat}/L_{\rm emit} \simeq 1$) with much higher superphoton resolution than in our fiducial simulations. The scattering quality factor increases with increasing number of superphotons as expected. The electron temperature varies slightly at small radius. The small magnitude of the variations, seen in all radiation variables as well as the observed spectra, suggests that underresolving scattering interactions may not compromise our radiation field solutions.

\section{Comparison with GRMHD and importance of cooling}
\label{app:coolcompare}

\begin{figure*}
\begin{tabular}{cc}
    \includegraphics[width=\columnwidth]{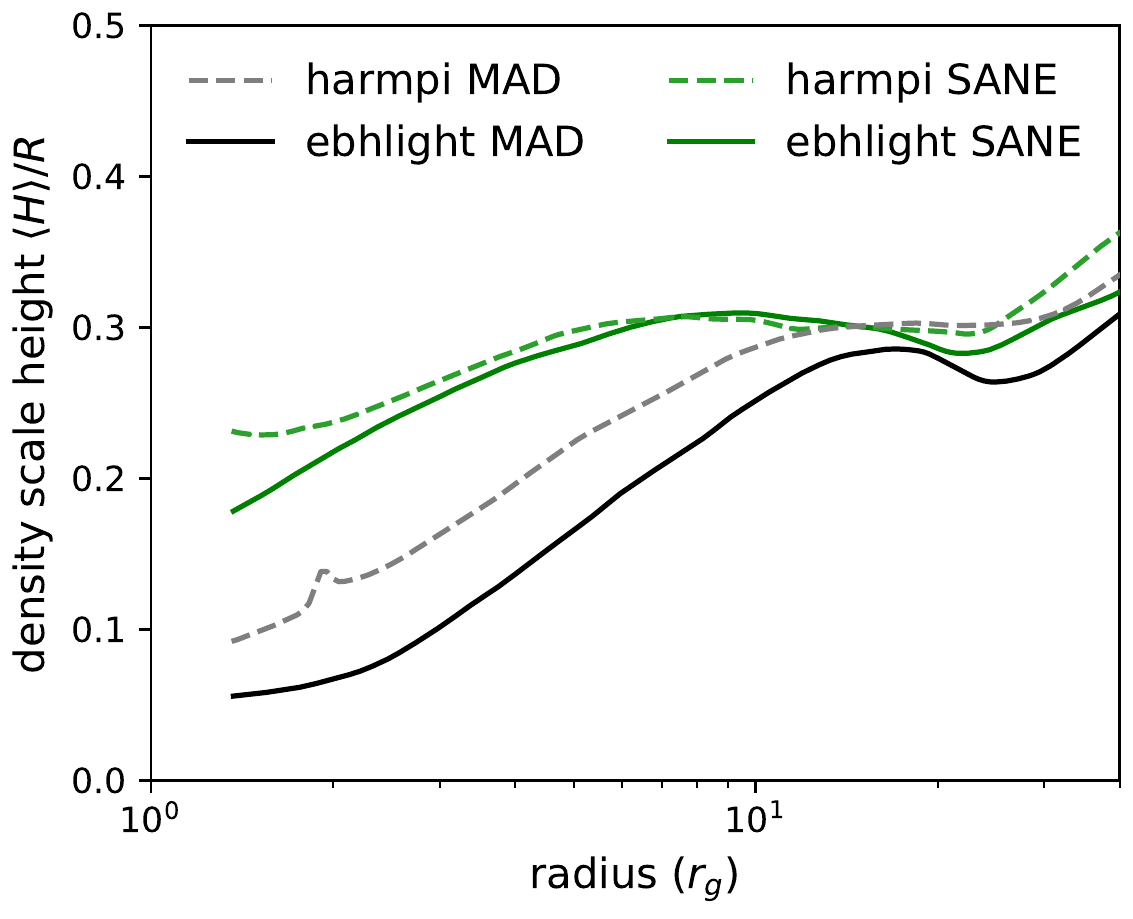} &
    \includegraphics[width=\columnwidth]{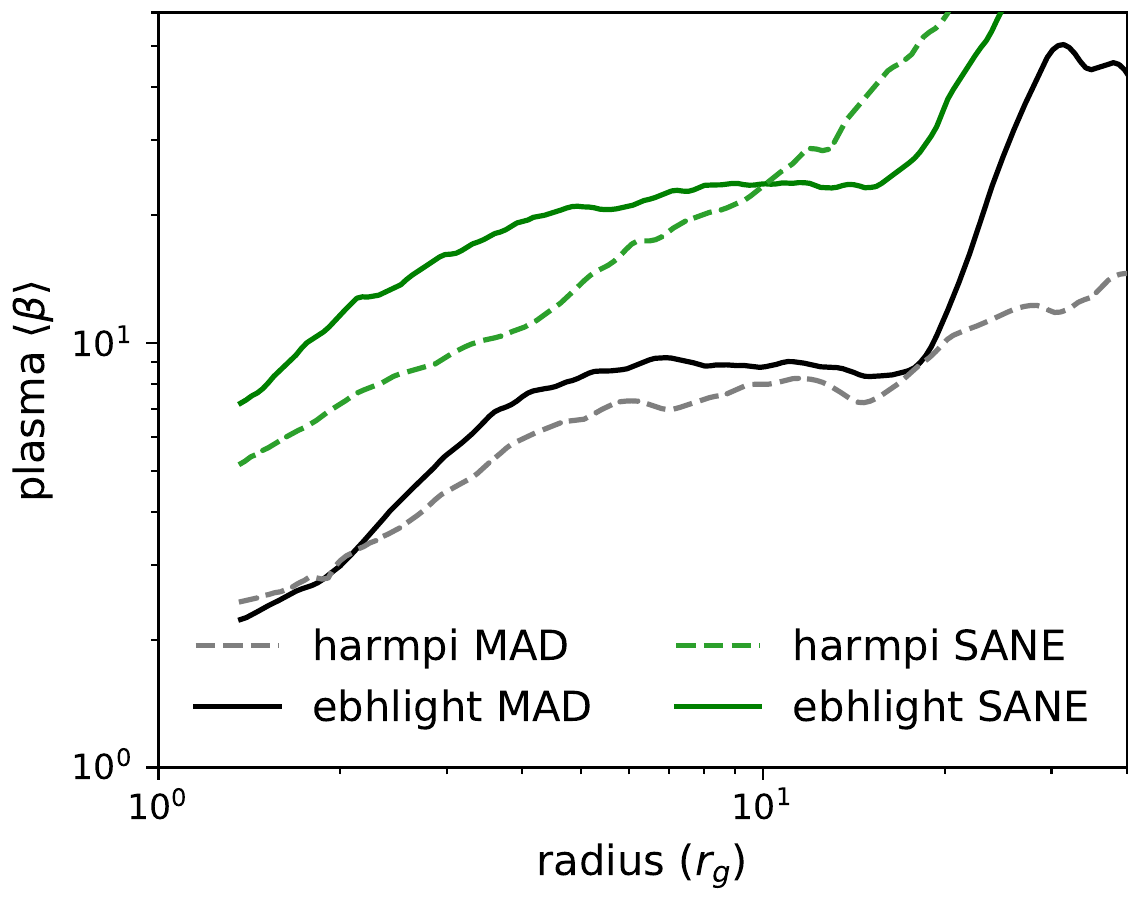}    
    \end{tabular}
    \caption{Comparison of radial profiles of $\langle H \rangle/R$ and plasma $\beta$ for MAD and SANE flow solutions obtained using the \textsc{ebhlight} and \textsc{harmpi} codes averaged from $5500-6500 \, r_g/c$.}
    \label{fig:sim_plots_harmpi}
\end{figure*}

\autoref{fig:sim_plots_harmpi} compares time averaged radial profiles of density scale height and plasma $\beta$ of SANE and MAD simulations obtained here with the W18 electron heating model and those from non-radiative simulations using the \textsc{harmpi} code\footnote{https://github.com/atchekho/harmpi} \citep{harmpiascl,dexter2020harmpi}. In both cases, the pairs of MAD and SANE simulations look similar. The density scale height from the \textsc{harmpi} calculation is somewhat higher than with \textsc{ebhlight}, while plasma $\beta$ is somewhat larger in the \textsc{ebhlight} SANE model than found with \textsc{harmpi}. Note that the initial conditions are not identical between these cases, and additionally the \textsc{ebhlight} simulations used an adiabatic index of $13/9$, while the \textsc{harmpi} simulations used $5/3$. The choice of adiabatic index is known to cause  differences in GRMHD calculations \citep[e.g.,][]{mignone2007,white2020}.

\begin{figure*}
    \begin{tabular}{cc}
    \includegraphics[width=\columnwidth]{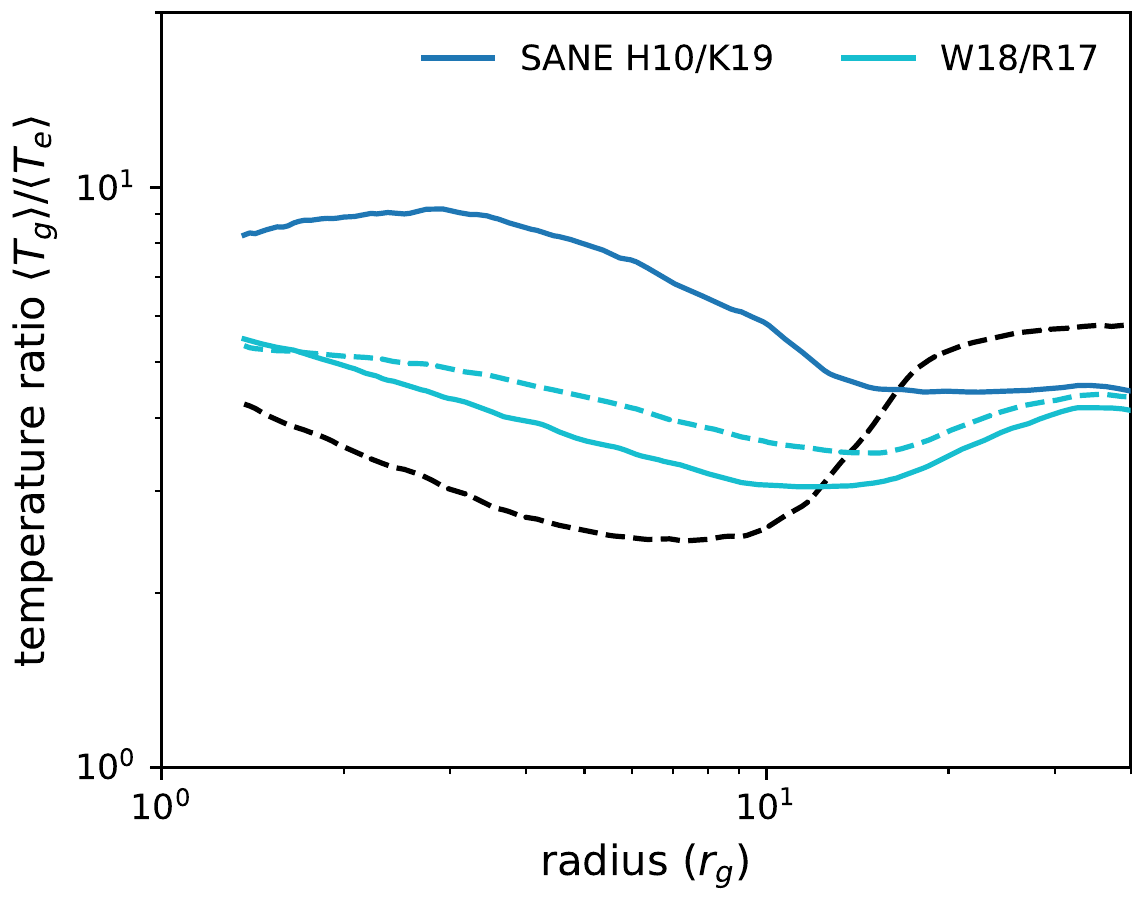} &
    \includegraphics[width=\columnwidth]{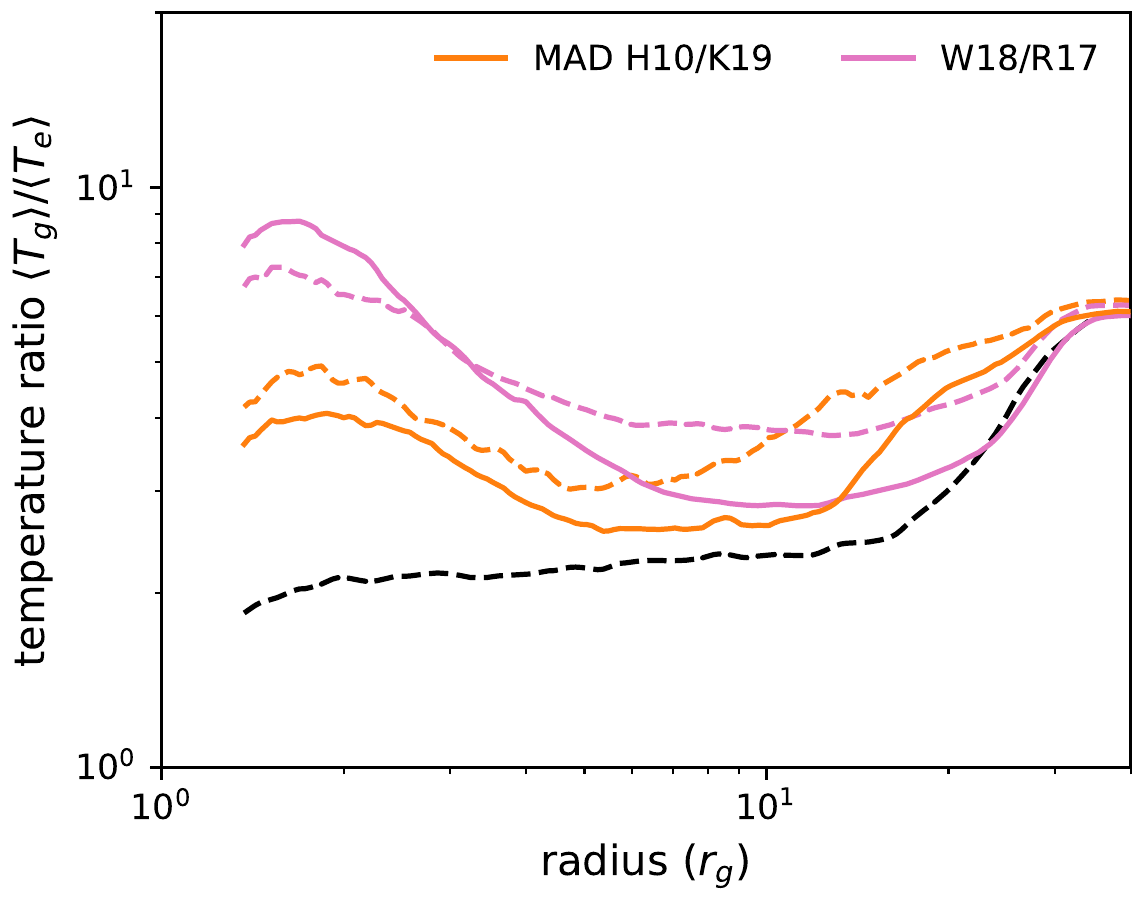}
    \end{tabular}
    \caption{Total fluid to electron temperature ratio for our SANE (left) and MAD (right) simulations. The dashed black lines show the initial state before the radiation field is turned on. Electron cooling is important in all models studied---the temperature ratio increases for $r \lesssim 10 \, r_g$ in all cases.}
    \label{fig:sim_plots_tpte}
\end{figure*}

\autoref{fig:sim_plots_tpte} compares the shell-averaged total fluid to electron temperature ratios for pure GRMHD (dashed black lines) to data from after the radiation is turned on (solid lines). The temperature ratio increases once radiative cooling is turned on, particularly at very small radii where most of the luminosity is emitted. The density-weighted temperature ratios are highest for the SANE H10/K19 and MAD W18/R17 models. Radiative cooling changes the electron temperatures near the horizon by factors of $\simeq 2-5$ on average.

\bsp    
\label{lastpage}
\end{document}